\documentclass[twocolumn,english,pra,showpacs,aps,superscriptaddress,amssymb,amsmath]{revtex4-1}

\usepackage{epsfig}
\usepackage{color}
\usepackage{comment}
\usepackage{appendix}

\usepackage[normalem]{ulem}
\usepackage[colorlinks=true,allcolors=blue,bookmarks=false,pdfusetitle]{hyperref}
\usepackage{url}

\usepackage{braket}

\newcommand{\sgn}{\text{sgn}}

\begin{document}

\title{Nonequilibrium quantum thermodynamics of determinantal many-body systems: Application to the Tonks-Girardeau and ideal Fermi gases}

\author{Y. Y. Atas}
\address{School of Mathematics and Physics, University of Queensland, Brisbane, Queensland 4072, Australia}
\address{Institute for Quantum Computing, Department of Physics and Astronomy, University of Waterloo, Waterloo N2L 3G1, Canada}
\author{A. Safavi-Naini}
\address{School of Mathematics and Physics, University of Queensland, Brisbane, Queensland 4072, Australia}
\address{ARC Centre of Excellence for Engineered Quantum Systems,
School of Mathematics and Physics, University of Queensland, Brisbane, QLD 4072, Australia} 
\author{K. V. Kheruntsyan}
\address{School of Mathematics and Physics, University of Queensland, Brisbane, Queensland 4072, Australia}

\date{\today}
\begin{abstract}
We develop a general approach for calculating the characteristic function of the work distribution of quantum many-body systems in a time-varying potential, whose many-body wave function can be cast in the Slater determinant form. Our results are applicable to a wide range of systems including an ideal gas of spinless fermions in one dimension (1D), the Tonks-Girardeau (TG) gas of hard-core bosons, as well as a 1D gas of hard-core anyons. In order to illustrate the utility of our approach, we focus on the TG gas confined to an arbitrary time-dependent trapping potential. In particular, we use the determinant representation of the many-body wave function to characterize the nonequilibrium thermodynamics of the TG gas and obtain exact and computationally tractable expressions---in terms of Fredholm determinants---for the mean work, the work probability distribution function, the nonadiabaticity parameter, and the Loschmidt amplitude. When applied to a harmonically trapped TG gas, our results for the mean work and the nonadiabaticity parameter reduce to those derived previously using an alternative approach. We next propose to use periodic modulation of the trap frequency in order to drive the system to highly non-equilibrium states by taking advantage of the phenomenon of parametric resonance. Under such driving protocol, the nonadiabaticity parameter may reach large values, which indicates a large amount of irreversible work being done on the system as compared to sudden quench protocols considered previously. This scenario is realizable in ultracold atom experiments, aiding fundamental understanding of all thermodynamic properties of the system.
\end{abstract}

\maketitle

\section{Introduction}

The study of thermalization in isolated quantum systems has fuelled the recent development of quantum thermodynamics  \cite{gemmer2009quantum,binder2018thermodynamics,Anders_Quantum_Thermodynamics,Kosloff_2013} and has highlighted its connection to the dynamics of quantum correlations, scrambling of quantum information, and entanglement dynamics~\cite{Ueda_Szilard_2011,Delgano_2014,Nandkishore_Many-B_2015,Dalessio2016,Jaramillo_2016}. Furthermore quantum thermodynamics plays an essential role in designing quantum heat engines. Here the unique quantum properties of the working fluid of the engine can be exploited to achieve quantum supremacy in terms of the engine efficiency and output power. In this work we consider a system amenable to analytical and numerical exploration in the context of quantum thermodynamics.

Recent advances in quantum simulation platforms utilizing a variety of atomic, molecular, and optical platforms have facilitated the observation of fundamental topics in quantum thermodynamics in the presence of different particle statistics, flexible dimensionality, and short-range and long-range interactions~\cite{Poletti_2015,Rey_MQC_2017,Schmiedmayer_ManyBody_2017, Friis2018, Brydges260, Monroe_Scrambling_2018}. 
More recently, a variety of experimental platforms, including trapped-ions and nitrogen vacancy (NV) centers, have observed quantum effects in the absence of interactions~\cite{SchmidtKaler2019, Eilon2019}. This underscores the need for theoretical studies which consider the role of quantum many-body interactions. The next generation of experiments will feature quantum many-body interactions, allowing one to use quantum correlations and  multi-particle entanglement as an additional resource to {control} 
the performance of quantum heat engines and quantum refrigerators. Trapped ultracold atomic gases lend themselves as a particularly promising platform in this respect \cite{Lutz_HO,Lutz_2016,Silva_2013,Chiara_2015,Poletti_2015,Jaramillo_2016,Del_Campo_2016_scaling,Li_2018,Andrei_2019,Gambassi_2019,Kurizki_2019,Arko_2020,fogarty2020many}  owing to a high degree of control over system parameters such as inter-atomic interactions and trapping potentials.

The quantum nature of the system complicates the calculation of physical quantities that are often considered in classical thermodynamics such as the mean work $\langle W(t)\rangle$. First, $W(t)$ is a randomly distributed quantity requiring two projective energy measurements at initial time  and at time $t$  \cite{Anders_Quantum_Thermodynamics,Kurchan,Talkner2007,Talkner_2012,Campisi2011_review}. Hence, quantum work is not represented by a Hermitian operator and hence cannot be regarded as an ordinary quantum observable \cite{Talkner2007} (see also Ref.~\cite{Work_generalised_measurement,cerisola2017using}).
Furthermore the outcome of these projective measurements depend on the transition probabilities between different quantum states.  This probabilistic nature implies that in order to fully characterize and understand the work performed during time $t$ one needs to know the full work probability distribution $P(W)$ or its corresponding characteristic function $G_\beta(\vartheta)$. Finally, since these calculations involve many-body expectation values, their computational complexity may grow exponentially with system size. This is a major obstacle for a theoretical optimization of the engine performance as it limits any such study to small systems.

{In this work we show that quantum many-body systems whose many-body wave functions can be represented as Slater determinants provide ideal platforms for exploring ideas of quantum thermodynamics. As we show here, we are able to recast the quantum thermodynamics of paradigmatic interacting many-body systems such as {a 1D gas of hard-core bosons and anyons} in terms of finite temperature dynamics, which itself can be reduced to the dynamics of single-particle wave functions. We illustrate this by using the Tonks-Girardeau (TG) gas \cite{Girardeau1960,Girardeau1965} which is realizable experimentally \cite{paredes2004tonks,Kinoshita2004,Naagerl2009,Weiss_2020}, is amenable to analytical treatment for its finite-temperature dynamics \cite{Atas2017a}, and as we have shown in our previous work~\cite{atas2019finite}, the system dynamics in a periodically modulated harmonic trap \cite{GangardtMinguzziExact,QuinnHaque2014,Atas2017a,Atas2017b,Scopa_2017,Scopa_2018} can be stable or unstable at a given modulation frequency depending on the amplitude of the modulation. }

In particular, we show that, for an arbitrary modulation of the trapping potential, thermodynamic quantities of interest, such as  the characteristic function $G_\beta(\vartheta)$, the moments $\langle W^n\rangle$ of the work distribution $P(W)$, and the nonadiabaticity parameter $Q^*(t)$, can all be evaluated using Fredholm determinants associated solely with single-particle wave functions. {This is similar to the functional determinant approach used in the treatment of time-dependent perturbations of Fermi gases \cite{Abanin_2005,dAmbrumenil_2005,Knap_2012,Schmidt_2018}, that relies on Levitov-Lesovik formula \cite{Levitov-Lesovik} for the characteristic function of charge distribution in electron transport, even though our derivations do not rely on the said formula.}

{We then apply our general results to study nonequilibrium thermodynamics of a periodically (sinusoidally) modulated harmonically trapped TG gas. Periodically driven systems are rather common in nature and form an important class of problems in quantum dynamics, which motivates our choice and distinguishes it from sudden quenches studied before. For periodic modulation of a harmonically trapped TG gas, we are able to present explicit analytical expressions for the aforementioned thermodynamic quantities, and we illustrate as an example that by simply tuning the amplitude of the modulation one can significantly increase the amount of energy pumped into the system (equivalent to the amount of work done on an isolated system) in a given time}. Our results can serve as a guide for designing quantum engine cycles based on periodic modulation rather than the more traditional sudden quenches (see, e.g., \cite{Jaramillo_2016}) {or other quantum control approaches} {such as shortcut to adiabaticity} \cite{deffner2014classical,guery2019shortcuts,Beau_2020,Busch_2020_adiabaticity}. Finally, our simple determinantal expression for the characteristic function allows us to explicitly calculate the expectation value $\langle e^{-\beta W} \rangle$ (where $\beta=1/k_BT$ is the inverse temperature) and hence demonstrate the validity of the quantum version of the Jarzynski equality \cite{Jarzynski_1997,Kurchan,Esposito_2009,Talkner_2012}, which relates quantum work associated with a nonequilibrium process with the equilibrium free energy difference of the initial and final states of this quantum many-body system.

\section{Thermodynamic quantities of determinantal many-body systems}
\label{sec:model}

{As an example of an interacting many-body system with determinantal structure of the quantum many-body function we consider the Tonks-Girardeau gas of $N$ bosons of mass $m$ interacting in one dimension via two-body hard-core interaction potential. Even though we are specifying the TG gas as the primary example for explicit illustration and discussion of our results, we note that all thermodynamic quantities calculated in this work are equally applicable to a 1D gas of spinless (or spin-polarized) noninteracting fermions \cite{Gong,Vicari2019,Free}, as well as to a 1D gas of hard-core anyons \cite{Girardeau_Anyons,Patu_Korepin_Anyons_2008,Patu}. In the purely bosonic or anyonic hard-core cases, the underlying statistics of the particles is governed by the constraints on the commutation relations of the field operators imposed by the hard-core diameter (with additional phase slips for anyons, compared to bosons), which at the same time allow for the Bose-Fermi or anyon-Fermi mapping to a pure ideal Fermi gas \cite{Girardeau1960,Girardeau1965,GirardeauWright2000,Pezer,yukalov2005fermi,Girardeau_Anyons,Patu_Korepin_Anyons_2008,Patu}. 
In all cases the evolution of the system in an external time-dependent one-body trapping potential $V(x,t)$ is governed by the free-fermion Hamiltonian
\begin{equation}
\hat{H}(t)=\sum_{j=1}^{N} \left[-\frac{\hbar^{2}}{2m} \frac{\partial^{2}}{\partial x_{j}^2}+V(x_{j},t)\right].
\label{Schrodinger_Eq}
\end{equation}
The TG or anyonic gas can equivalently be described as the limiting case of the Lieb-Lininger model with infinitely strong two-body $\delta$-function interaction potential \cite{Lieb-Liniger1963}.

The  thermodynamics scenarios that we consider correspond to the system prepared at time $t=0$ in a thermal equilibrium with a reservoir at temperature $T$ and chemical potential $\mu$. It is then decoupled from the reservoir and evolves unitarily in an arbitrary time-dependent trapping potential $V(x,t)$. According to Bose-Fermi or anyon-Fermi mapping, the evolving many-body wave function of the system can be expressed in terms of the fermionic many-body wave function 
\begin{equation}
\Psi_{s}(x_{1},...,x_{N};t)\!=\!A(x_{1},...,x_{N})\Psi_{s}^{F}(x_{1},...,x_{N};t), 
\label{FB-mapping}
\end{equation}
where $A(x_{1},...,x_{N})$ is a statistical factor that ensures the right symmetry of the wave function under particle exchange, whereas the pure fermionic wavefucntion $\Psi_{s}^{F}(x_{1},\dots,x_{N};t)$ can be written as Slater determinant of the single-particle wave functions $\phi_{s_{i}}(x,t)$, where $\lbrace s_{j} \rbrace_{j=1,2,\dots,N}$ are a set of relevant quantum numbers.
For a TG gas of hard-core bosons, $A(x_{1},...,x_{N})$ is a unit antisymmetric function given by  $A(x_{1},\ldots,x_{N}) = \prod_{1\leq j \leq i \leq N}
\sgn(x_{i} - x_{j})$, with $\sgn(0)=0$, whereas for hard-core anyones it has an additional phase factor \cite{Girardeau_Anyons,Patu_Korepin_Anyons_2008,Patu}. The relevant property that ultimately makes the calculation of our thermodynamic quantities identical to those of free fermions is that $|A(x_{1},...,x_{N})|^2=1$ and hence can be dropped from the final results; for the cases where $|A(x_{1},...,x_{N})|^2$ returns $0$, this corresponds to particles sharing the same position and is hence already taken into account by the fermionic Slater determinant which respects the Pauli exclusion principle.

With these introductory remarks in mind, we now turn to the discussion of thermodynamic work performed on the system by the external trapping potential $V(x,t)$. We can evaluate the quantum work} using two projective energy measurements, one performed at time $t=0$ and the other at time $t$. The work in each realization of the experiment is given by the difference in the measured energies, and the associated work probability distribution function in an ensemble of realizations is given by \cite{Kurchan,Talkner2007,Talkner_2012,Campisi2011_review,Jarzynski_2015}
\begin{align}
 P(W,t)\!=\!\sum_{N=0}^{\infty}\!\sum_{s}\!\sum_{{s'}} p_{N,s}^{(0)}\,p_{{s'}|s}^{(t)}\, \delta(W\!-\!E^{(t)}_{N,{s'}}\!+\!E_{N,s}^{(0)}). 
\label{work_distribution}
\end{align}
{In the following, we will explicitly omit the time dependence of the work distribution and conveniently write $P(W)\equiv P(W,t)$}. 
Here, $p_{N,s}^{(0)}$ represents the probability of the first measurement returning an energy eigenvalue $E_{N,s}^{(0)}$ corresponding to the $N$-particle eigenstate $\vert \Psi_{s}(0)\rangle$ (where we omit the index $N$ for notational simplicity) of the initial Hamiltonian $\hat{H}(t\!=\!0)$, satisfying $\hat{H}(t\!=\!0)\vert \Psi_{s}(0)\rangle=E_{N,s}^{(0)}\vert \Psi_{s}(0) \rangle$.
We consider initial thermal equilibrium states described by the grand-canonical ensemble, and therefore $p_{N,s}^{(0)}$ is given by the normalized Gibbs factor, $p_{N,s}^{(0)}=\frac{1}{\mathcal{Z}_{0}}e^{-\beta(E_{N,s}^{(0)}-\mu N)}$. Here, $\beta=1/k_BT$ is the initial inverse temperature (with $k_B$ the Boltzmann constant),  $\mu$ is the initial chemical potential, and $\mathcal{Z}_0$ is the corresponding grand-canonical partition function. The system is then isolated from the reservoir and undergoes unitary evolution generated by $\hat U(t)=\mathcal{T}e^{-i \int_0^t dt' \hat{H}(t')/\hbar}$. The resulting state is $\vert \Psi_{s}(t)\rangle=\hat{U}(t)\vert\Psi_{s}(0) \rangle$.
Next, a second projective energy measurement at time $t$ returns one of the instantaneous energy eigenvalues $E_{N,s'}^{(t)}$, with the corresponding instantaneous eigenstate denoted via $\vert \Phi_{s'}^{(t)}\rangle$, such that $\hat H(t) \vert \Phi_{s'}^{(t)}\rangle = E^{(t)}_{N,{s'}} \vert \Phi_{s'}^{(t)}\rangle$. Therefore, the second probability entering into Eq.~\eqref{work_distribution}, $p_{s'\vert s}^{(t)}$, is given by the transition probability from  $\vert \Psi_{s}(t)\rangle$ to $\vert \Phi_{s'}^{(t)}\rangle$, namely  $p_{s'\vert s}^{(t)}=\vert \braket{\Phi_{{s'}}^{(t)}|\Psi_{s}(t)}\vert^{2}=\vert \braket{\Phi_{{s'}}^{(t)}|\hat U(t) \Psi_{s}(0)}\vert^{2}$. Finally, the $\delta$-function in Eq.~\eqref{work_distribution} ensures the conservation of energy, such that in each realization of the protocol the work is given by $W(t)=E_{N,s'}^{(t)}-E_{N,s}^{(0)}$.

However, in practice, it is often more convenient to work with the characteristic function of the work distribution, which is given by the Fourier transform of $P(W)$,
\begin{equation}
G_{\beta}(\vartheta)=\int \mathrm{d}W e^{i \vartheta W/\hbar}P(W) =\langle e^{i \vartheta W/\hbar}\rangle. 
\label{Fourier_Transform_Work}
\end{equation}
The characteristic function is the generating function of the moments of the distribution of work through successive differentiation,
\begin{equation}
\langle W(t)^{n} \rangle = (-i \hbar)^{n } \frac{d^{n}G_{\beta}(\vartheta)}{d\vartheta^{n}}\Big{|}_{\vartheta=0},
\label{moments}
\end{equation}
with the first moment corresponding to the mean work $\langle W(t) \rangle$.

{Evaluating $G_\beta(\theta)$ for a generic quantum many-body system is a daunting task. 
However,  for systems with the determinantal structure of the many-body wave-function as in Eq.~\eqref{FB-mapping}, one can find numerically tractable expressions for $G_\beta(\theta)$ for a general $V(x,t)$. Since the mapping \eqref{FB-mapping} holds true for the instantaneous eigenfunctions $\Phi_s^{(t)}(x_{1},...,x_{N})$ as well, the determinantal structure of the fermionic many-body wave function reduces the evaluation of Eq.~\eqref{Fourier_Transform_Work}} to that of evaluating integrals involving time-evolved single-particle wave functions $\phi_{s_i}(x,t)$ and the instantaneous single-particle eigenfunctions $\phi_{s_i}^{(t)}(x)$ with respective instantaneous eigenenergies $E_{s_i}^{(t)}$. Here, $\phi_{s_{i}}(x,0)$ are the energy eigenstates for the initial trapping potential $V(x,0)$ with eigenenergies $E_{s_{i}}^{(0)}$, such that the total energy of the  $N$-particle system system is given by $E_{N,s}^{(0)}=\sum_{i=1}^{N}E_{s_{i}}^{(0)}$. 

Using the properties of Fredholm integral equations and determinants~\cite{lenard1966one,bornemann2010numerical}, the characteristic function can be written as (see Appendix~\ref{F} for details),
\begin{equation}
G_{\beta}(\vartheta)=\frac{\mathcal{D}_{\hat{K}}^{k}(\vartheta)}{\mathcal{D}_{\hat{F}_0}^{f_{0}}}=
\frac{\mathrm{det}(1+\hat{K})}{\mathrm{det}(1+\hat{F}_0)},
\label{eq:Gcompact}
\end{equation}
where the superscripts $k$ and $f_0$ indicate the kernels of the respective Fredholm determinants, $\mathcal{D}_{\hat{K}}^{k}$ and $\mathcal{D}_{\hat{F}_0}^{f_{0}}$, given by
\begin{align}
\label{k_kernel_main_text}k(x,y)&= \sum_{i,j}\phi_{i}(x,t) k_{ij}(t) \left(\phi_{j}^{(t)}(y)\right)^*,\\
\label{f0_kernel_main_text} f_0(x,y)&=z \sum_{i} e^{-\beta E_{i}^{(0)}}\phi_{i}(x,0)\left(\phi_{i}^{(0)}(y)\right)^*,
\end{align}
with
\begin{equation}k_{ij}(t)\!=\!z e^{-E_{i}^{(0)} \left(\beta+i \vartheta /\hbar\right) + i \vartheta E_{j}^{(t)}/\hbar} \! \int \!dw \phi_{i}^*(\omega, t) \phi_{j}^{(t)}(\omega),
\end{equation}
and  $z\!=\!e^{\beta \mu}$ denoting the fugacity. Here, the operators $\hat{K}$ and $\hat{F_{0}}$ are integral operators with kernels given by $k(x,y)$ and $f_{0}(x,y)$, respectively. For instance, the action of $\hat{K}$ on an arbitrary function $\xi(r)$ is given by $(\hat{K}\xi)(w)=\int_{\mathbb{R}} k(w,v)\xi(v)dv$.

Equation~\eqref{eq:Gcompact} is a significant result of this work as it allows us to express the characteristic function of the work distribution function of the quantum many-body system exclusively in terms of the single-particle quantities. This can be made more explicit if one rewrites Eq.~\eqref{eq:Gcompact} in terms of the eigenvalues of $\hat K$ and $\hat F_0$, which we denote by $\lbrace \Lambda_{i}^{(\hat{K})} \rbrace_{i=0,1,\dots}$ and $\lbrace \Lambda_{i}^{(\hat{F}_0)} \rbrace_{i=0,1,\dots}$, respectively. The eigenvalues for the operator $\hat K$ (and similarly for $\hat F_0$) and their corresponding eigenfunctions $\lbrace \theta_{i}^{(\hat K)} (w)\rbrace_{i=0,1,\dots}$ satisfy the equation 
\begin{equation}
\int_{\mathbb{R}}dv ~k(w,v)\theta_{i}^{(\hat{K})}(v)=\Lambda_{i}^{(\hat{K})}\theta_{i}^{(\hat{K})}(w),\label{eigenvalue_equation}
\end{equation}
with a similar equation for the operator $\hat F_0$, where $k(w,v)$ is replaced by $f_0(w,v)$. 
Using the expansion of the determinant of an operator as a product over its eigenvalues, Eq.~(\ref{eq:Gcompact}) can be rewritten as
\begin{equation}
G_{\beta}(\vartheta)=\prod_{i}\left( \frac{1+\Lambda_{i}^{(\hat{K})}}{1+\Lambda_{i}^{(\hat{F}_0)}}\right),\label{eigenvalue_expansion_G}
\end{equation}
which expresses the characteristic function of the work distribution of the TG gas under an arbitrary driving protocol as an infinite product over the eigenvalues of two integral operators with kernels given by (\ref{k_kernel_main_text}) and (\ref{f0_kernel_main_text}).

While the exact analytical evaluation of the eigenvalues is not always possible, {the Fredholm determinant representation of the characteristic function, Eq.~\eqref{eq:Gcompact},} offers a compact and a computationally practical way to efficiently evaluate many-body thermodynamic quantities of interest numerically, in terms of single-particle wave functions. For sufficiently smooth kernels, a simple tabulation of the kernel on a finite grid and the use of Nystr\"om classical quadrature routine enables one to numerically evaluate the Fredholm determinant with small absolute errors \cite{bornemann2010numerical}. In Appendix \ref{AppendixB}, we show that the eigenvalues $\Lambda_{i}^{(\hat K)}$ (and $\Lambda_{i}^{(\hat F_0)}$ ) {and the determinant} of the Fredholm integral equation can {also} be efficiently obtained through direct diagonalization of a matrix with elements proportional to the overlap between the time evolved and instantaneous eigenfunctions.

The above discussion is written generally for a thermal initial state. However, in many quantum simulation platforms, the relevant initial state is a pure state such as the many-body ground state of the initial Hamiltonian $\vert \Psi_0(0)\rangle$. In this regime, the sensitivity of the system to the driving protocol can be probed via the Loschmidt echo, whose amplitude is equivalent to the zero-temperature characteristic function $\mathcal G(t)=G_{\beta\to \infty}(t)$ \cite{Silva_2008}. The Loschmidt echo amplitude is given by (see, e.g., Ref.~\cite{Loschmidt-review} for a review)
\begin{equation}
\mathcal G(t)=\langle \Psi_0(0) \vert e^{i \hat H(0) t/\hbar} \mathcal{T}e^{-i \int_0^t dt' \hat H(t') /\hbar}\vert \Psi_0(0)\rangle, 
\end{equation}
with the Loschmidt echo itself given by 
\begin{equation}
L(t)=\vert \mathcal G(t)\vert ^2. 
\end{equation}

The Loschmidt echo gives the {survival probability} of the initial eigenstate $\vert \Psi_0(0)\rangle$ first evolved forward in time according to $\hat{H}(0)$ and then backward in time according to the dynamics generated by $\hat{H}(t)$. 
Its utility is in characterizing the survival of a quantum state when an imperfect time-reversal is applied. Since $\hat{H}(0) \vert\Psi_0(0)\rangle = E_0 \vert \Psi_0(0)\rangle $ and $\mathcal{T}e^{-i \int_0^t dt'\hat H(t') /\hbar}\vert \Psi_0(0)\rangle=\hat{U}(t)\vert \Psi_0(0)\rangle= \vert \Psi_0(t) \rangle$ is the time-evolved state of the system, one can rewrite the 
Loschmidt echo as 
\begin{equation}
L(t)=\vert \langle \Psi_0(0) | \Psi_0(t) \rangle\vert ^2.\label{L_overlap}
\end{equation}
We now recognize this expression as the dynamical fidelity, $\mathcal{F}(t)=\vert \langle \Psi_0(0) | \Psi_0(t) \rangle\vert ^2$, which is simply the squared overlap between the initial state and the time-evolved state of the system.

In order to derive a compact and computationally tractable expression for the  Loschmidt echo $L(t)$, we use a similar procedure to that used in the derivation of Eq.~\eqref{eq:Gcompact} (see Appendix~\ref{L}). The echo amplitude $\mathcal{G}(t)$ can then be written as the determinant of an $N\times N$ matrix $\mathbf{A}$ (where $N$ is the number of particles in the system) containing the overlaps between the initial and time evolved single-particle eigenfunctions $\phi_{n}(x,t)$ [see Eq.~\eqref{Solution_single_schrodinger} below], and hence 
\begin{equation}
L(t)=|\mathrm{det}\mathbf{A}(t)|^{2},
\end{equation}
where 
\begin{equation}
A_{m n}(t)=\int_{-\infty}^{\infty}\phi_{m}^{\ast}(x,0)\phi_{n}(x,t) \,dx, \label{Overlap_Matrix}
\end{equation}
and $m,n=0,1,\dots,N-1$.
This determinantal result for the Loschmidt echo reproduces the expression derived previously {for the TG gas in Refs.~\cite{delcampo2011,L_1D_Bose_Gas,lelas2012pinning,Pons2012}. An equivalent  determinantal expression for a system of 1D free fermionic chains has recently been derived in Ref.~\cite{Caux_2020}.}

\section{Tonks-Girardeau gas in a harmonic trap}

In this section, we apply the Fredholm determinant formalism to a time-varying harmonic potential $V(x,t)=m\omega^2(t)x^2/2$. This model offers the advantage of being analytically solvable, in addition to being typical of ultracold atom experiments, and thus provides an ideal platform for studying the work distribution and other thermodynamic quantities of interest. 
We point out that the thermodynamic quantities for the TG in a harmonic trap with arbitrary time-variation of the trap frequency $\omega(t)$ were previously derived by Jaramillo \textit{et al.} in Ref.~\cite{Jaramillo_2016} using the scale invariance of the corresponding many-body wave function. 
However, our determinantal approach allows us to derive a more general result for the mean work $\langle W(t) \rangle$ and the nonadiabaticity parameter $Q^*(t)$ that is valid for an arbitrary spatial shape of $V(x,t)$ beyond the simple harmonic trapping. An immediate application of our general results to this limit---as is done in this section---reproduces their results for the harmonic trapping.

Our numerical results in the harmonic trapping regime are further distinguished from previous studies due to the use of a nontrivial driving protocol wherein the trap potential is modulated sinusoidally in time\, with $V(x,t)=m\omega^2(t)x^2/2$ and $\omega(t)^2=\omega(0)^2[1-\alpha \sin (\Omega t)]$. Here, $\Omega$ is the modulation frequency and $\alpha$ characterizes the modulation amplitude. Under such periodic modulation, the TG trap displays a rich phase diagram in the $(\Omega,\alpha)$ parameter space \cite{atas2019finite}, including regions of stable (bounded) and unstable (exponentially growing due to parametric resonance) dynamics \cite{QuinnHaque2014}. The latter regime is particularly advantageous for creating highly nonequilibrium states, with very large values of the nonadiabaticity parameter $Q^*(t)\gg 1$. This in turn corresponds to large amount of work that can be done on the system accompanied with modest changes in the trap size, which may prove useful in optimising the performance of a quantum refrigerator with the TG gas serving as the working medium.

With an arbitrary time-varying harmonic potential, the single-particle Schr\"odinger equation is exactly solvable and the time evolved single-particle wave functions $\phi_{n}(x, t)$ can be obtained through a simple scaling transformation \cite{PerelomovBook,GangardtMinguzziExact} given by
\begin{equation}
\phi_{n}(x,t)=\frac{1}{\sqrt{\lambda}}\phi_{n}\left(\frac{x}{\lambda},0\right)\exp\left[ i \frac{mx^{2}}{2\hbar}\frac{\dot{\lambda}}{\lambda}-i\frac{\mathcal{E}_{n}(t)\,t}{\hbar}\right]. \label{Solution_single_schrodinger}
\end{equation}
Here, the initial wave functions $\phi_{n}(x,0)$ are the Hermite-Gauss polynomials of a quantum mechanical $1$D harmonic oscillator,
\begin{equation} 
\phi_{n}(x,0)=\frac{\exp\left(-x^{2}/2l^{2}_{\mathrm{ho}}(0)\right)H_{n}(x/l_{\mathrm{ho}}(0))}{\pi^{1/4}\sqrt{2^{n}n!l_{\mathrm{ho}}(0)}},
\label{Gauss_Hermite_eigenstate}
\end{equation}
with frequency $\omega(0)$,  energy eigenvalues $E_{n}^{(0)}=\hbar \omega(0)(n+1/2)$, and harmonic oscillator length $l_{\mathrm{ho}}(0)=\sqrt{\hbar/m\omega(0)}$.
Furthermore, the scaling function $\lambda(t)$ is a solution to the Ermakov-Pinney equation \cite{ermakov,atas2019finite}, \begin{equation}
\ddot{\lambda}(t)+\omega^{2}(t)\lambda(t)=\frac{\omega(0)^{2}}{\lambda^{3}(t)}, \label{ermakov}
\end{equation}
with initial conditions $\lambda(0)$=1 and $\dot{\lambda}(0)=0$, whereas the time-dependent phase factor $\mathcal{E}_n(t)$ in Eq.~\eqref{Solution_single_schrodinger} is defined in terms of  $\lambda(t)$ and reads as
\begin{equation}
\mathcal{E}_n(t)=\hbar \omega(0)\left(n+\tfrac{1}{2}\right) \,\frac{1}{t} \int_0^t \frac{dt'}{\lambda^2(t')}.
\label{Et}
\end{equation}

\subsection{The characteristic function for a thermal state}

We proceed by evaluating the characteristic function $G_{\beta}(\vartheta)$ in Eq.~\eqref{eigenvalue_expansion_G} for the general case of a thermal initial state at inverse temperature $\beta$.
The details of the derivation for the eigenvalues of the Fredholm integral equation involving Hermite-Gauss polynomials can be found in Appendix \ref{G}. Here we report only the final results. 
Following the substitution of (\ref{partition_function_eigenvalue}) and (\ref{lambda_K}) in Eq.~(\ref{eigenvalue_expansion_G}), the characteristic function of the work distribution takes the form
\begin{equation}
G_{\beta}(\vartheta)=\prod_{i=0}^{\infty} \frac{1+z\xi^{i+1/2}}{1+z e^{-\beta \hbar \omega(0)(i+1/2)}},
\label{exact_harmonic_G}
\end{equation}
where $z\xi^{i+1/2}$ are the eigenvalues $\Lambda_i^{\hat K}$ of the Fredholm integral equation with kernel (\ref{k_kernel}). In Appendix~\ref{G} [see Eq.~\eqref{xi_eq}] we provide an explicit analytic expression for $\xi$ which contains the dependence on $\vartheta$. We note that the denominator in Eq.~\eqref{exact_harmonic_G} can be recognized as the equilibrium partition function of free fermions in a harmonic trap with frequency $\omega(0)$. For $\vartheta=0$,  we have $\xi(\vartheta=0)=e^{-\hbar\beta \omega(0)}$ and hence $G_{\beta}(0)=1$, which can be directly seen from the definition, Eq.~(\ref{Fourier_Transform_Work}).

\subsection{The mean work}

We can now use the above results to compute the mean work performed in the system during the driving protocol. Differentiating the logarithm of $G_{\beta}(\vartheta)$ with respect to $\vartheta$ and using the fact that $G_{\beta}(0)=1$, we find $G_{\beta}^{\prime}(\vartheta=0)=(\log  G_{\beta}(\vartheta))^{\prime}|_{\vartheta=0}$. Thus, using Eq.~\eqref{exact_harmonic_G}, we arrive at the expression for the mean work,
\begin{align}
\nonumber\langle W(t) \rangle &= -i \hbar \,\frac{dG_{\beta}(\vartheta)}{d\vartheta}\Big{|}_{\vartheta=0}\\
\nonumber&= \left(\frac{\omega(t)}{\omega(0)}\zeta(t)-1 \right) \\
&\hskip 15pt \times \hbar \omega(0)\sum_{i}\frac{z(i+1/2)e^{-\beta\hbar \omega(0)(i+1/2)}}{1+ze^{-\beta\hbar \omega(0)(i+1/2)}},
\label{avW}
\end{align}
where the time dependent parameter $\zeta(t)$ is given by (from Appendix~\ref{G})
\begin{equation}
    \zeta(t)=\frac{1}{2\omega(0)\omega(t)}\left( \dot{\lambda}(t)^{2}+\frac{\omega^{2}(0)}{\lambda(t)^2}+\omega^{2}(t)\lambda(t)\right).
    \label{zeta}
\end{equation}

The last line in Eq.~(\ref{avW}) corresponds to the initial thermal average energy of the system at inverse temperature $\beta$, i.e., $\notag \langle \hat{H}(0)\rangle=-\partial \log \mathcal{Z}_{0}/\partial \beta$. Therefore, the mean work performed in the system after a driving time $t$ is given by
\begin{equation}
 \langle W(t) \rangle=   \left(\frac{\omega(t)}{\omega(0)}\zeta(t)-1 \right)\langle \hat{H}(0)\rangle,
 \label{average_work}
\end{equation}
 and is thus proportional to the initial equilibrium internal energy of the gas $\langle \hat{H}(0) \rangle$, with a (time-dependent) constant of proportionality given by the term in the brackets.

This result provides an explicit analytic expression that allows one to calculate the mean work $\langle W(t) \rangle$ for a wide range of Hamiltonian parameters. Furthermore, it provides a computationally tractable expression for evaluating the mean work done on or by the system during a time $t$. The only explicit unknown here is the solution to a simple second-order ODE, the Ermakov-Pinney equation \eqref{ermakov}, for the scaling parameter $\lambda(t)$. It is worth pointing out, however, that this particular ODE for a sinusoidally modulated trap affords analytic solutions in term of Mathieu's functions \cite{atas2019finite}, or else can be easily solved numerically.

\begin{figure}[tbp]
\includegraphics[width=8.6cm]{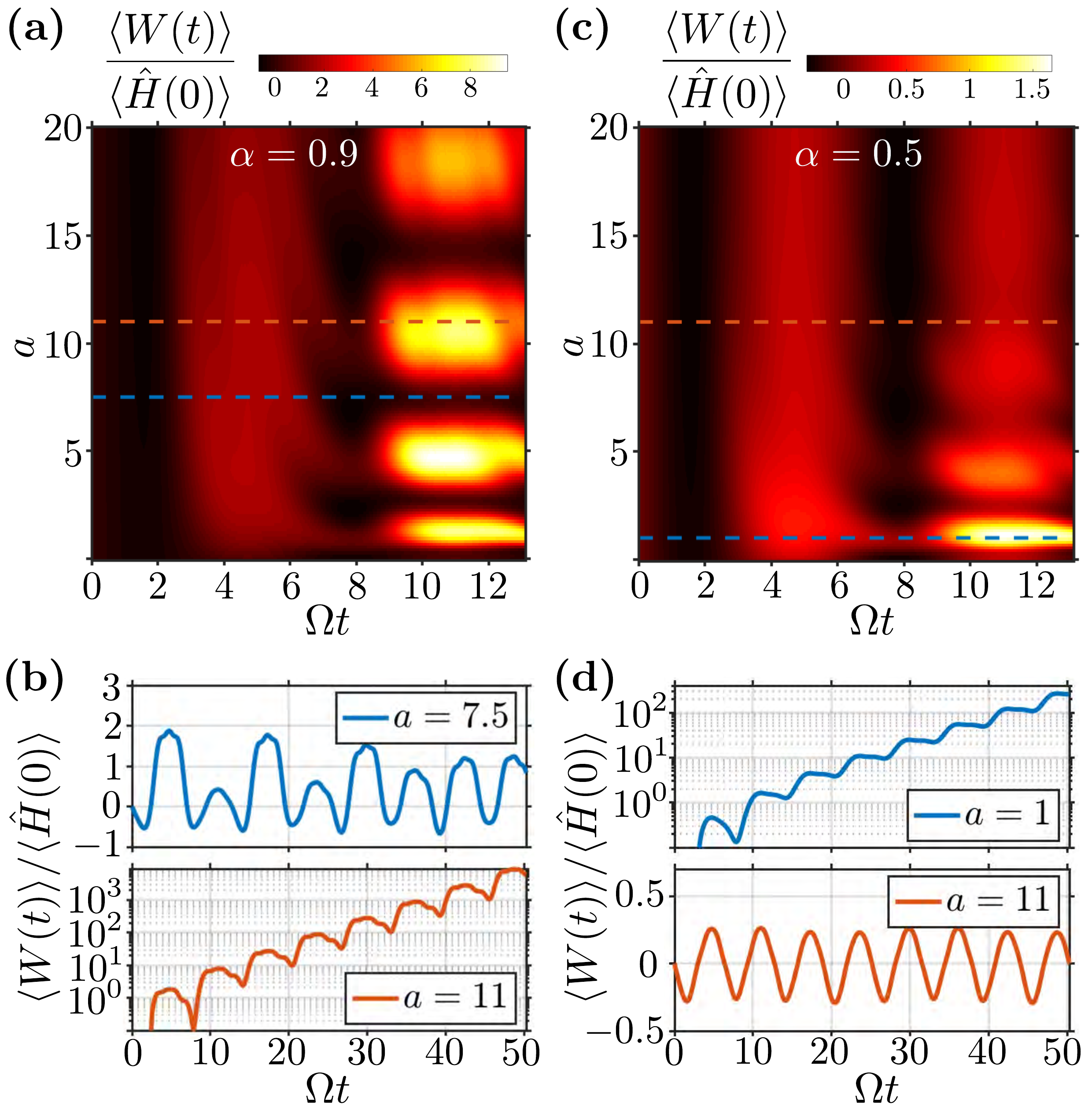}
\caption{
The mean work $\langle W(t)\rangle$ of a driven Tonks-Girardeau gas in a harmonic trap, normalized by the initial thermal equilibrium energy $\langle \hat{H}(0)\rangle$. The driving protocol used here is periodic modulation of the trap strength $V(x,t)=m\omega(t)^2x^2/2$, with $\omega(t)^2=\omega(0)^2[1-\alpha \sin (\Omega t)]$. Panel (a) shows  $\langle W(t) \rangle$ as a function of the dimensionless time $\Omega t$ and the driving frequency parameter $a\!\equiv \![2\omega(0)/\Omega]^2$, for the driving amplitude $\alpha\!=\!0.9$, whereas panel (c) is for $\alpha\!=\!0.5$. This parametrization is the same as the one used in the stability diagram of Fig.~3 of Ref.~\cite{atas2019finite}. Fixing the value of $a$ (at given $\alpha$) corresponds to a specific realization of the driving protocol. Panels (b) and (d) show, respectively, the cuts of $\langle W(t)\rangle$ at constant $a$ from panels (a) and (c), but for a longer time span.  The examples of $\langle W(t)\rangle$ in (b) at $a=7.5$ and $a=11$ correspond, respectively, to stable (semi-periodic) and unstable (exponentially growing) dynamics invoked by the sinusoidal drive at this value of the modulation amplitude $\alpha=0.9$. Similar examples of unstable behavior can be realized at other values of the parameter $a$ from the high-intensity horizontal bands in panel (a), corresponding to parametric resonances at values of $a$ slightly above $a\!=\!j^2$, where $j\!=\!1,2,3,...$ ~\cite{atas2019finite}. For comparison, in the examples of panel (d), which are for a smaller value of the modulation amplitude ($\alpha=0.5)$, the curve for $a=11$ is no longer unstable, and as an unstable example we show $\langle W(t)\rangle$ at the primary parametric resonance $a=1$.
}
\label{fig:W}
\end{figure}

In Fig.~\ref{fig:W} we show the mean work $\langle W(t) \rangle$ done on or by a TG gas in a periodically modulated trap $V(x,t)=m\omega(t)^2x^2/2$, with $\omega(t)^2=\omega(0)^2[1-\alpha \sin (\Omega t)]$, as a function of time and the dimensionless driving frequency parameter $a\!\equiv \![2\omega(0)/\Omega]^2$, for two values of the modulation amplitude $\alpha$. For relatively large values of $\alpha$ (which we note are bounded between $0 \leq \alpha \leq 1$), such as in Fig.~\ref{fig:W}\,(a) with $\alpha=0.9$, we see several high intensity horizontal bands emerging dynamically with time. These finite-width bands correspond to the values of $a$ that lie within the unstable regions of the stability phase diagram of the system \cite{atas2019finite} occurring  around and predominantly slightly above $a=1, 4, 9, 16, \dots$, i.e., around integer ratios of $2\omega(0)/\Omega$. In these unstable regions, the energy of the system grows exponentially with time  due to the phenomenon of parametric resonance. In this driving scenario, with an explicit example shown in Fig.~\ref{fig:W}\,(b) for $a=11$, the mean work done \emph{on} the system ($\langle W(t) \rangle >0$) can reach very large values $\langle W(t)\rangle /\langle \hat{H}(0)\rangle  \gg 1$, even though the relative change in the instantaneous frequency $\omega(t)$ of the trap at time $t$, i.e., at the end of any particular realization of the driving protocol, is not wildly different from $\omega(0)$; somewhat counter-intuitively, the final value of $\omega(t)$ can even be smaller than $\omega(0)$, which under an adiabatic drive would have to correspond to work done \emph{by} the system under adiabatic expansion, corresponding to $\langle W(t) \rangle <0$. In contrast, for values of $a$ outside the parametric resonance band, the dynamics is stable, and $\langle W(t)\rangle$ is bounded and generally quasiperiodic. An example of such stable dynamics is shown in panel (b) for $a=7.5$. We note that $\langle W(t) \rangle$ can oscillate between positive and negative values. 

For smaller amplitudes of the drive, the widths of the parametric resonance bands along $a$ become narrower and only the lower-order resonances get efficiently excited (see Fig.~\ref{fig:W}\,(c)). In Fig.~\ref{fig:W}\,(d) we show resonantly growing $\langle W(t)\rangle /\langle \hat{H}(0)\rangle  \gg 1$ for the lowest order resonance ($a=1$) and $\alpha=0.5$, which is similar to the previous example of $a=11$ at $\alpha=0.9$. However, the dynamics for $a=11$ at $\alpha=0.5$ is no longer unstable and we see nearly sinusoidal modulation $\langle W(t) \rangle $ between negative and positive values. In fact this is the behavior that is expected for nearly adiabatic drive, during which the mean work alternates between being done \emph{on} the system or \emph{by} the system, following respectively the opening or tightening the trap as per modulation of its frequency $\omega(t)$. We confirm this below using the nonadiabaticity parameter $Q^*(t)$, shown in Fig.~\ref{fig:Q} below.

\begin{figure}[tbp]
\includegraphics[width=8.5cm]{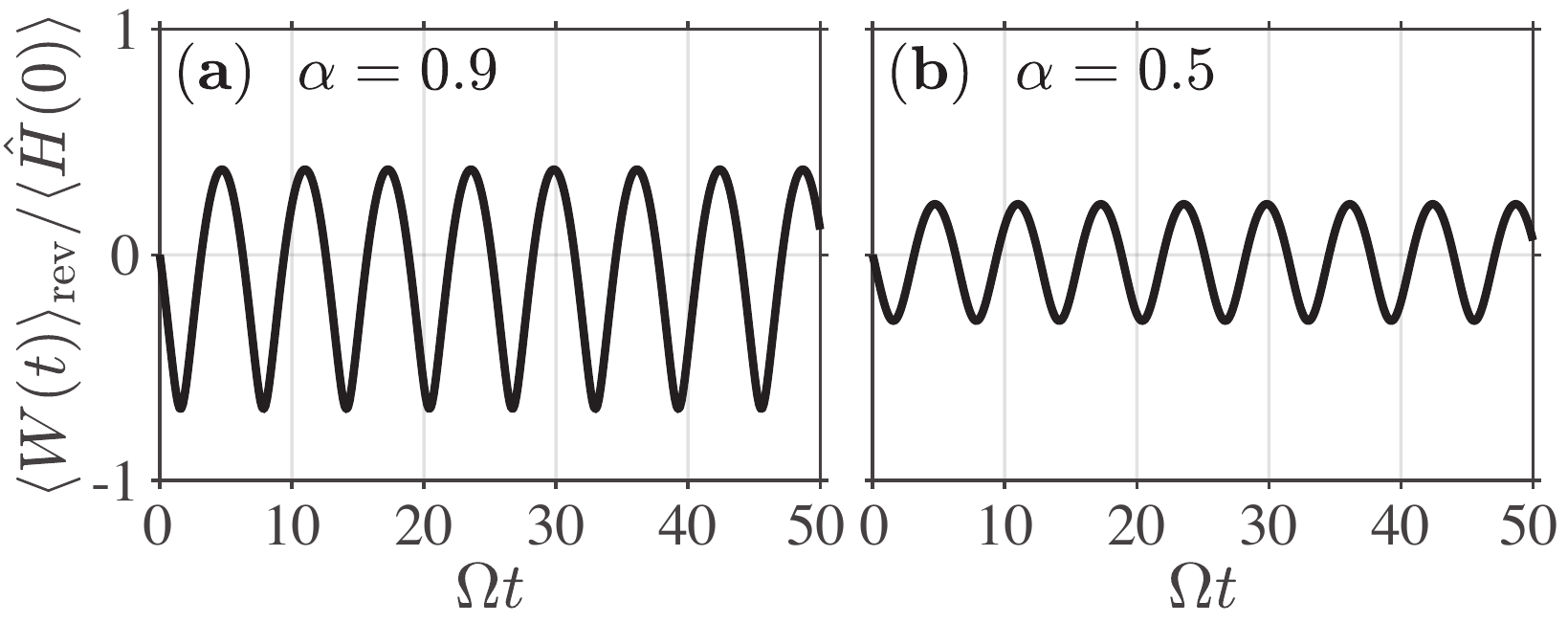}
\caption{{The mean reversible work $\langle W(t)\rangle_{\mathrm{rev}}$ for the same driving protocol as in Fig.~\ref{fig:W}, for two values of the driving amplitude  $\alpha$. The difference between the mean work $\langle W(t)\rangle$ shown in Fig.~\ref{fig:W} and the reversible work shown here is the irreversible work.}
}
\label{fig:W_rev}
\end{figure}

{As we show below, the parameter $\zeta(t)$ in Eq.~\eqref{average_work}  is equivalent to the nonadibaticity parameter $Q^*(t)$ that we introduce in Eq.~\eqref{Q_param}. Accordingly, the mean work calculated with the nonadiabaticity parameter set to unity, $\zeta(t)=Q^*(t)=1$, corresponds to the reversible or adiabatic work, $\langle W(t) \rangle_{\mathrm{rev}}=\langle W(t) \rangle \vert_{Q^*=1}$, whereas the difference between the total work and the reversible work represents the irreversible part, $\langle W(t) \rangle_{\mathrm{irrev}}=\langle W(t) \rangle-\langle W(t) \rangle_{\mathrm{rev}}$.
For the periodic trap modulation as in Fig.~\ref{fig:W}, one can show that the reversible work $\langle W(t) \rangle_{\mathrm{rev}}=\big(\omega(t)/\omega(0)-1\big)\langle \hat{H}(0)\rangle=\big( \sqrt{1-\alpha \sin(\Omega t)}-1\big)\langle \hat{H}(0)\rangle $, which is shown in Fig.~\ref{fig:W_rev} for the same two values of the driving amplitude $\alpha$ as in Fig.~\ref{fig:W}; it is independent of the ratio of the driving frequency $\Omega$ and the initial trap frequency $\omega(0)$, as expected. Indeed, under the adiabatic drive, the system follows the trap changes adiabatically irrespectively of the initial trap frequency.
}

To summarize, the examples of Fig.~\ref{fig:W} illustrate the rich variety of driving protocols that can be realized in a periodically modulated TG gas in a harmonic trap. This variety stems from the rich stability phase diagram of the system which contains regions of parametrically resonant unstable dynamics.  Since the behavior is fully determined by two parameters, namely the driving amplitude and the driving frequency, these examples highlight the utility of a periodically modulated TG gas as the working fluid of a quantum machine. For instance it may be used to perform large irreversible work without the need for large changes in the trap size, governed by $\alpha$ and the modulation frequency $\Omega$.

\subsection{Work probability distribution}

\begin{figure}[tbp]
\includegraphics[width=7.8cm]{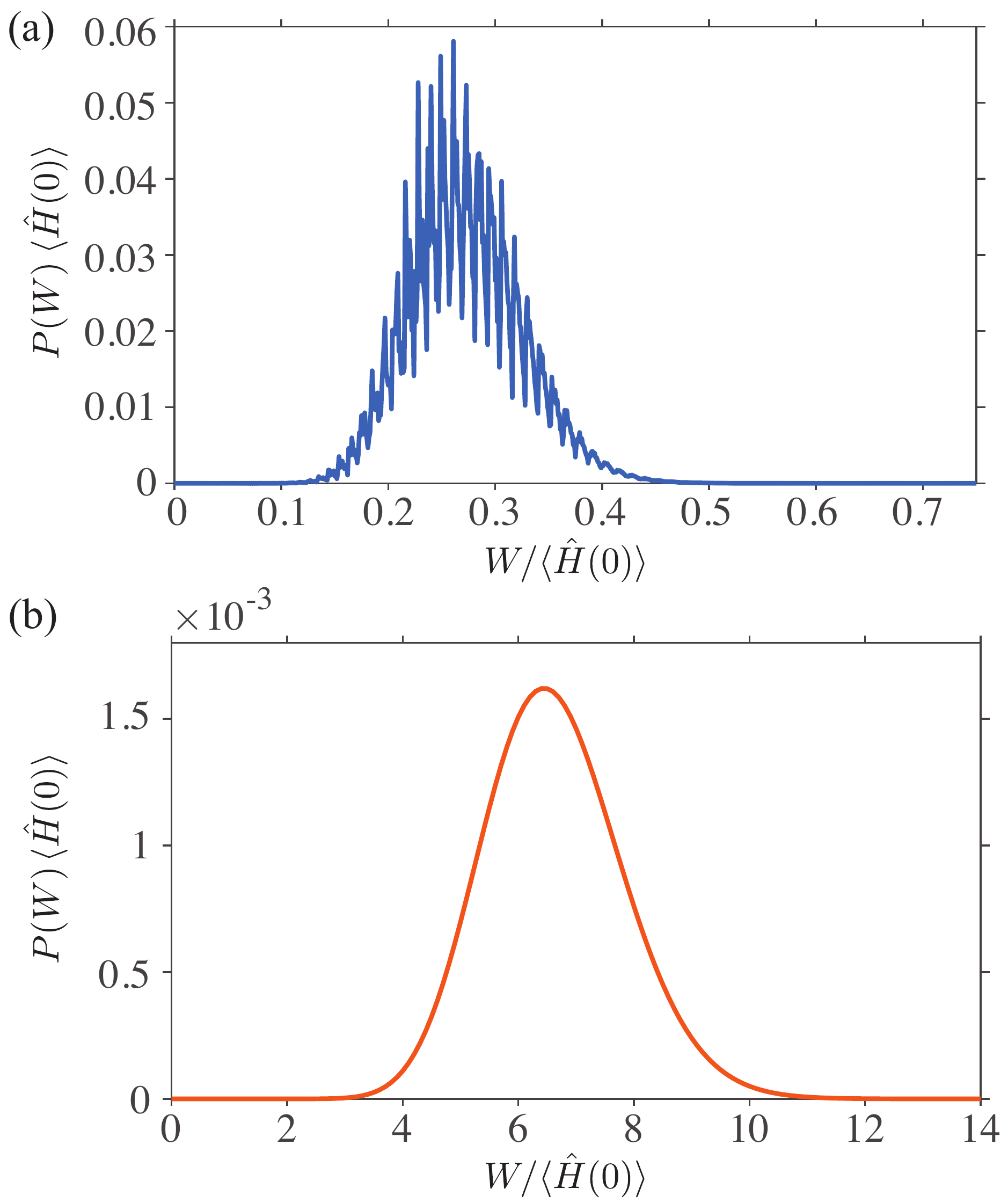}
\caption{
The work probability distribution function $P(W)$ versus $W$ for a TG gas in a periodically modulated harmonic trap as in Fig.~\ref{fig:W}, evaluated at time $\Omega t=10$. Panels (a) and (b) are, respectively, for the same parameters as the two curves in Fig.~~\ref{fig:W}\,(b), representing examples of stable and unstable dynamics. The chemical potential $\mu$ is chosen to result in the average number of particles $\langle N \rangle=20$, and $\langle \hat{H}(0) \rangle$ is evaluated at the temperature well below the temperature of quantum degeneracy, $k_BT/\langle N\rangle \hbar \omega(0)=0.1$.
}
\label{fig:P_W}
\end{figure}

Given the exact analytic expression for the characteristic function of the work distribution, Eq.~\eqref{exact_harmonic_G}, we can evaluate not only the mean work $\langle W(t) \rangle$ of a driven TG gas, but also any higher-order moments of the work probability distribution, or indeed the full probability distribution $P(W)$ by taking the inverse Fourier transform of Eq.~\eqref{Fourier_Transform_Work} at a particular time instance $t$. In Fig.~\ref{fig:P_W}(a) and (b) we show representative examples of $P(W)$ under the same driving protocol as in Fig.~\ref{fig:W}, evaluated at time $\Omega t=10$ for unstable and stable dynamics, respectively.

In the stable regime (see Fig.~\ref{fig:P_W}(a)) the probability distribution $P(W)$ is relatively narrow and is localized around a relatively small values of  $ W/\langle \hat{H}(0)\rangle$. In this case, the transition probabilities $p_{s\vert s'}$ in Eq.~\eqref{work_distribution} involve transitions to instantaneous eigenstates whose eigenenergies are not far away from the initial eigenenergies. Therefore, the resulting transition probabilities strongly depend on the structure of the overlaps between the particular eigenstates involved. Accordingly, $P(W)$ in this example displays a fine structure that reflects the discrete nature of energy levels $E^{(0)}_{N,s}$ and $E^{(t)}_{N,s'}$, that contribute to the random outcomes of the measurement of $W=E^{(t)}_{N,s'}-E^{(0)}_{N,s}$ due to sensitivity to the various overlaps. In contrast, in the unstable regime (see Fig.~\ref{fig:P_W}(b)) the distribution is broad, smooth, and centered around relatively large values of $W/\langle \hat{H}(0)\rangle$. In this case, the discrete nature of energy levels washes out as the typical instantaneous energies $E^{(t)}_{N,s'}$ are far from the initial eigenenergies, and all corresponding overlaps are small and similar in magnitude.

\subsection{The nonadiabaticity parameter}

If the drive between the initial and final states is nonadiabatic in the quantum sense \cite{adiabatic}, then $\phi_{n}(x,t)$ will not be an eigenstate of the instantaneous Hamiltonian $\hat{H}(t)$. 
To quantify the degree of nonadiabaticity of the driving protocol, we introduce the nonadiabaticity parameter
\begin{equation}
    Q^*(t)= \frac{\langle \hat{H}(t) \rangle}{\langle \hat{H}(t)\rangle_{\rm ad}}.  \label{Q_param}
\end{equation}
This parameter is the ratio of the average energy measured with the actual protocol and the average energy obtained through an adiabatic driving. Clearly, $Q^*(t)\geq 1$ with $Q^*(t)=1$ when the evolution is adiabatic. {This parameter has the same meaning as the one introduced by Husimi \cite{Husimi1953} and used in the case of a single-particle quantum harmonic oscillator \cite{Lutz_HO}.}

For the TG gas in a harmonic trap with an arbitrary time dependence of $\omega(t)$, the adiabatic driving energy is related to the initial energy via the relation
\begin{equation}
\langle \hat{H}(t) \rangle_{\rm ad}= \frac{\omega(t)}{\omega(0)}\langle \hat{H}(0) \rangle,
\label{H_ad}
\end{equation}
which can be rewritten as $\langle \hat{H}(t) \rangle_{\rm ad}\!=\! \langle \hat{H}(0) \rangle /\lambda^{2}_{\rm ad}(t)$, where $\lambda_{\rm ad}(t)\!=\![\omega(0)/\omega(t)]^{1/2}$ is the solution to the Ermakov-Pinney equation (\ref{ermakov}) in the adiabatic limit $\ddot{\lambda}\!\approx \!0$. Equation \eqref{H_ad} follows from the fact that $\langle \hat{H}(0) \rangle\!=\!\sum_{N,s} p_{N,s}^{(0)}\langle \Psi_s(0)\vert \hat{H}(0) \vert \Psi_s(0) \rangle \!=\! \sum_{N,s}p_{N,s}^{(0)} E_{N,s}^{(0)}$ with $E_{N,s}^{(0)}\!=\!\hbar \omega(0)\sum_{i=1}^{N} (s_{i}+1/2)$, and that the adiabatic mean energy is given by $\langle \hat{H}(t) \rangle_{\rm{ad}}\!=\! \sum_{N,s}p_{N,s}^{(0)} E_{N,s}^{(t)}$, with $E_{N,s}^{(t)}\!=\!\hbar \omega(t)\sum_{i=1}^{N} (s_{i}+1/2)$. 

We now use Eqs.~\eqref{Q_param} and \eqref{H_ad} to express $\langle \hat H (t)\rangle$ in terms of $\langle \hat H(0)\rangle$, $Q^*(t)$, $\omega(t)$, and $\omega(0)$. This allows us to rewrite the expression for the mean work as, 
\begin{equation}
     \langle W(t) \rangle = \left[\frac{\omega(t)}{\omega(0)}Q^*(t)-1\right] \langle \hat{H}(0) \rangle. 
     \label{averageWgeneral}
\end{equation}

Thus, the knowledge of  $\langle W(t) \rangle$ is equivalent to the knowledge of the nonadiabaticity parameter $Q^*(t)$ and vice versa. Indeed, by comparing Eq.~(\ref{averageWgeneral}) with Eq.~(\ref{average_work}), we identify the nonadiabaticity parameter $Q^*(t)$ with $\zeta(t)$,
 \begin{equation}
     Q^*(t)=\zeta(t),\label{Q_final_thermal}
 \end{equation}
where $\zeta(t)$ itself is given by Eq.~\eqref{zeta}.

\begin{figure}[tbp]
\includegraphics[width=8.6cm]{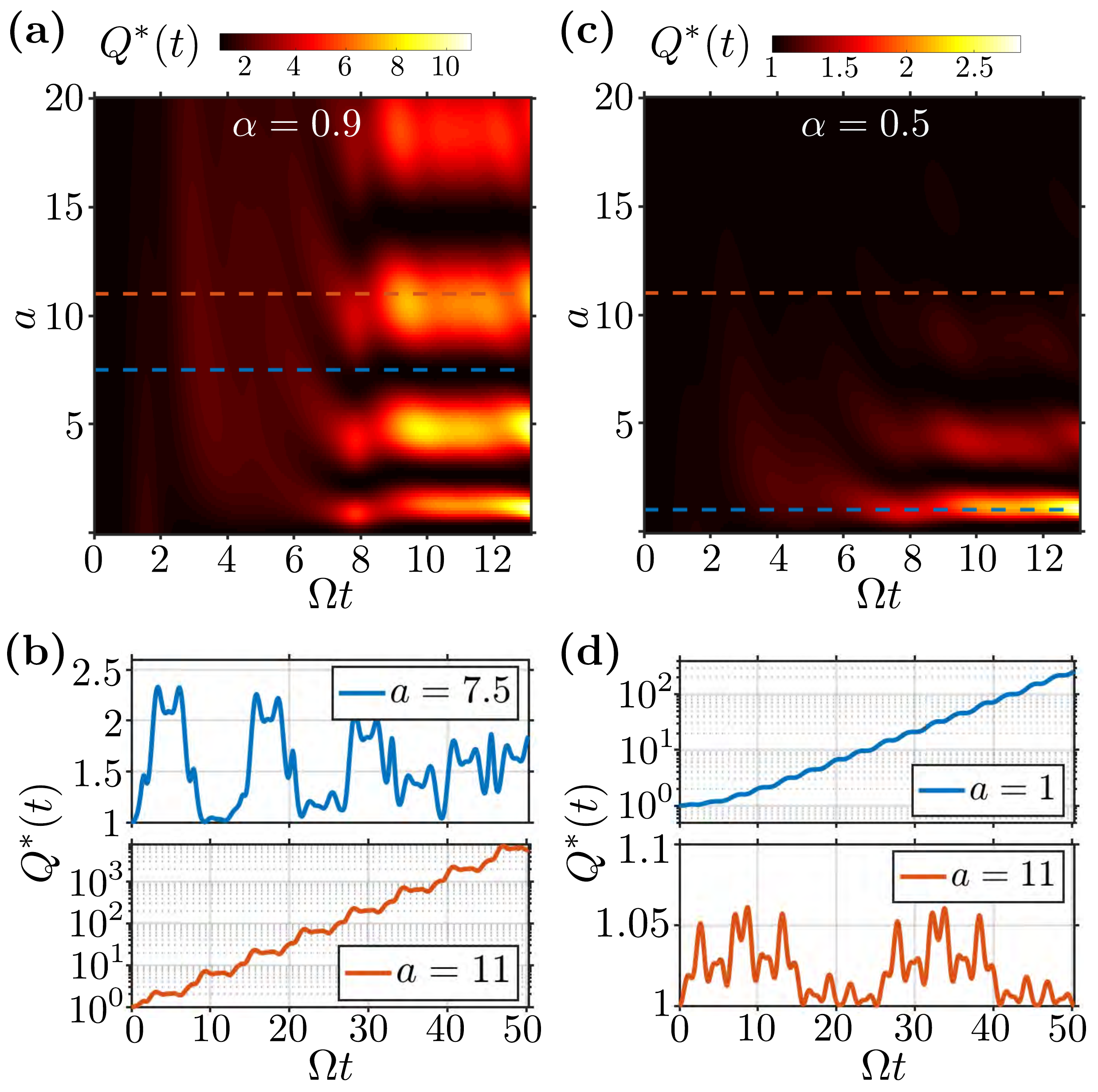}
\caption{The nonadiabaticity parameter $Q^*(t)$ for the Tonks-Girardeau gas in a periodically modulated harmonic trap. All parameters and representative examples are the same as in Fig.~\ref{fig:W}. 
}
\label{fig:Q}
\end{figure}

In Fig.~\ref{fig:Q} we show the nonadiabaticity parameter $Q^*(t)$ for a TG gas in a periodically modulated trap $V(x,t)=m\omega(t)^2x^2/2$, with $\omega(t)^2=\omega(0)^2[1-\alpha \sin (\Omega t)]$, as a function of time and the driving frequency parameter $a\!\equiv \![2\omega(0)/\Omega]^2$, for the same values of $\alpha$ as in Fig.~\ref{fig:W}, i.e., $\alpha=0.9$ for the panels (a) and (b) on the left, and $\alpha=0.5$ for the panels (c) and (d) on the right. The examples of cuts at $a=11$ in panel (b) and $a=1$ in panel (d) are from the unstable region. In this regime $Q^*(t)$ reaches values much greater than one, corresponding to the driving protocol generating highly non-equilibrium final states. In contrast, at $a=11$ in panel (c), the system evolves under nearly adiabatic dynamics where $Q^*(t)\approx 1$ for all time. Finally, at $a=7.5$ in panel (b), we observe stable (quasiperiodic) dynamics, but with intermediate values of the nonadiabaticity parameter $Q^*(t)$.

\subsection{The nonadiabaticity parameter for pure states}

It is also instructive to consider driving the system starting from a pure state rather than a thermal state. Since the dynamics induced by the modulation of the trap frequency is governed, in both cases, by single-particle dynamics and by a single scaling parameter $\lambda(t)$, one expects to find the same $Q^*(t)$ as the one for the thermal initial state.  For a pure initial state, the expectation values appearing in Eq.~(\ref{Q_param}) can be computed easily.  Lets us first consider a pure state $\ket{\Psi_{s}}$ in ket notation, characterized by $N$ non-negative integers $s=s_{1},s_{2},\dots,s_{N}$. The choice of the set of integers $\{s_{i}\}$ is quite arbitrary, corresponding in general to the ground or excited many-body states, but we will see that the final result for $Q^*(t)$ is independent of this choice. The time-evolved wave function for a pure state at time $t$ is obtained from the Slater determinant of single-particle time evolved wave-functions,
\begin{equation}
     \Psi^F_{s}(x_{1},\dots,x_{N},t)=\frac{1}{\sqrt{N!}}\,\underset{1\leq n\leq m\leq N}{ \mathrm{det}} \phi_{s_{n}}(x_{m},t),
\end{equation}
and has mean energy that is a simple sum of the respective single-particle energy eigenvalues $E_{s_n}$,
\begin{equation}
    \langle \hat{H}(t) \rangle=\sum_{n=1}^{N}E_{s_{n}}(t).
\end{equation}
with $\hat H(t) \phi_{s_n}(x,t) \!=\! E_{s_n}\phi_{s_n}(x,t)$ and hence  $\int dx\, \phi_{s_n}^*(x,t) \hat H(t) \phi_{s_n}(x,t)\!=\!E_{s_n}(t)$. Using Eq.~\eqref{Solution_single_schrodinger} for the time-evolved eigenfunctions, we find
\begin{equation}
E_{s_n}(t)=\hbar \bar \omega(t)(s_n+1/2),
\end{equation} 
where we have defined
\begin{equation}
\bar{\omega}(t)\equiv \frac{1}{2\omega(0)}\left( \dot{\lambda}^{2}(t)+\omega^{2}(t)\lambda^2(t)+\frac{\omega^{2}(0)}{\lambda^{2}(t)}\right).
\label{bar_omega}
\end{equation}

On the other hand, the adiabatic expectation value appearing in the denominator of Eq.~(\ref{Q_param}) is given by 
\begin{equation}
    \langle \hat{H}(t) \rangle_{ad}=\sum_{n=1}^{N}E_{s_{n}}^{(t)}=\sum_{n=1}^{N}\hbar \omega(t)\left(s_{n}+\frac{1}{2}\right).
\end{equation}

Taking the ratio of these two expectation values of the Hamiltonian for evaluating $Q^*(t)$, we see that the dependence on the specific configuration $s$ cancels out, and we obtain
\begin{equation}
\label{eq:Qpure}
Q^*(t)= \frac{\overline{\omega}(t)}{\omega(t)}.
\end{equation}
With $\bar{\omega}(t)$ given by Eq.~\eqref{bar_omega}, we see that Eq.~\eqref{eq:Qpure} is exactly the same as the one derived previously for the general case of a thermal initial state, Eq.~(\ref{Q_final_thermal}). We further emphasize that the equivalence of the results for $Q^*(t)$ for a pure and thermal initial states is enabled only by the scale invariance of the underlying many-body dynamics, which was exploited in Ref.~\cite{Jaramillo_2016,Beau_2020} to derive the same expression for the nonadiabaticity parameter using an alternative method.
{For the same scale-invariance reasons, rather than relying on the determinantal structure of the many-body state, equivalent expressions to the nonaidiabaticity parameter and the mean work derived here have recently been reported for a strongly interacting, unitary Fermi gas in three dimensions  \cite{Deng_2018} and the harmonic Calogero–Sutherland model \cite{Beau_2020} to which the TG and ideal Fermi gases are distinct limiting cases.}

\subsection{Loschmidt echo}

We now turn to the discussion of the Loschmidt echo in a harmonically trapped TG gas. We remind the reader that while all the results below are valid for an arbitrary time dependence of the trap frequency $\omega(t)$, the numerical examples will be given for the periodic modulation, consistent with the material presented earlier in this section. 
The exact expression of the Loschmidt echo amplitude, for an initial pure state with fixed number of particles $N$ in the system, can be written as (see Appendix \ref{L} for details), 
\begin{equation}
\mathcal{G}(t)=\left(\frac{2}{\widetilde{\lambda}(t) \lambda(t)}\right)^{N^{2}/2}e^{ i \left(E_n^{(0)}-\mathcal{E}_n(t) \right)tN^2/2\hbar},
\label{L-ampl}
\end{equation}
where $E_n^{(0)}=\hbar \omega(0)\left(n+\frac{1}{2}\right)$ and with
\begin{equation}
\widetilde{\lambda}(t)\equiv 1+\frac{1}{\lambda^2(t)}-i\frac{\dot{\lambda}(t)}{\lambda(t) \omega(0)}.
\label{tildelambda}
\end{equation}

The Loschmidt echo itself is therefore given by 
\begin{equation}
L(t)=\Big{|} \frac{2}{\widetilde{\lambda}(t) \lambda(t)}\Big{|}^{N^{2}}=l(t)^{N^2}. 
\label{Loschmidt}
\end{equation}

\begin{figure}[tbp]
\includegraphics[width=8.6cm]{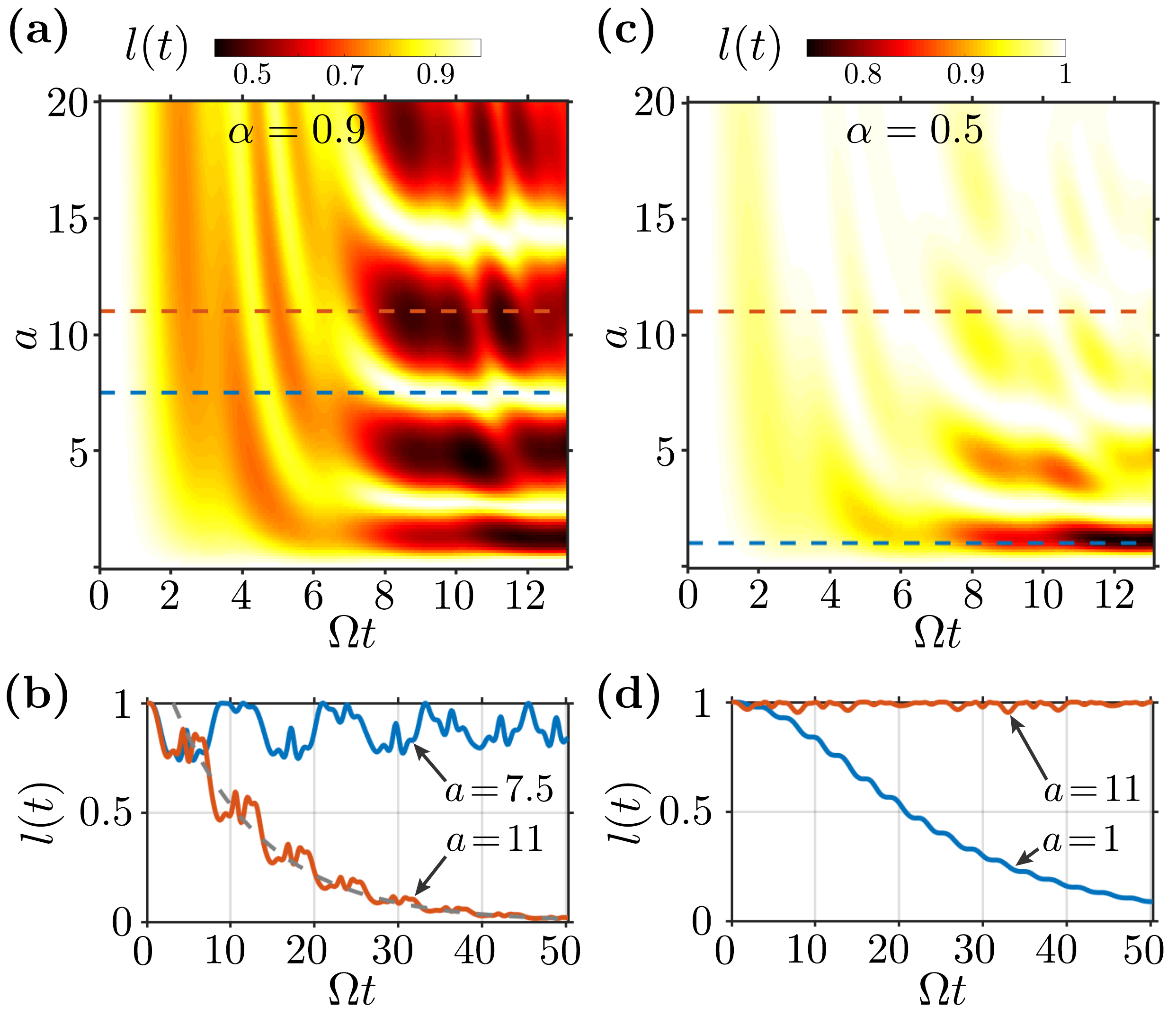}
\caption{The base $l(t)$ of the Loschmidt echo $L(t)=l(t)^{N^2}$ for the Tonks-Girardeau gas in a periodically modulated harmonic trap. All parameters and representative examples are the same as in Figs.~\ref{fig:W} and \ref{fig:Q}. The gray (dashed) line in panel (b) is the exponentially decaying prediction of $l(t)=e^{\pi \gamma}e^{-i \gamma \tau}$ (where $\tau=\Omega t$), with the dimensionless decay rate $\gamma=|\mathrm{Im}(\nu)|/2$, where $\nu$ is the Floquet exponent \cite{atas2019finite} equal to $\nu=1-0.182531i$ in this example. For the respective Poincare maps, explaining the typical features seen in these Loschmidt echo curves, see Fig.~\ref{fig:P} and text.
}
\label{fig:L}
\end{figure}

The Loschmidt echo, or the dynamical fidelity between the initial and final states,  characterizes the survival or recurrence probability of the initial state after the system has evolved for time $t$ under the specific driving protocol. Under periodic modulation of the trap frequency, the dynamics of the Loschmidt echo can be used to diagnose the stability of the TG gas for a given set of parameters. In Fig.~\ref{fig:L}, we contrast the behavior of $L(t)$ in the stable and unstable regimes, for the same parameters as in Figs.~\ref{fig:W} and \ref{fig:Q}. 
In the stable regime, the Loschmidt echo possesses a semi-periodic behavior since Mathieu's function (see Ref.~\cite{atas2019finite} for the mapping between the solutions to the Ermakov-Pinney equation for the scaling parameter and Mathieu's function) in this region is bounded with a real and non-integer Floquet exponent $\nu$. On the other hand, in the unstable regime the Loschmidt echo displays an exponential decay. In this region of the parameter space, the Floquet exponent $\nu$ is complex. The scaling function $\lambda(t)$ 
then contains an exponentially decaying envelope, in addition to its semi-periodic behavior \cite{atas2019finite}. Hence we observe an overall exponential decay [see the grey dashed line in Fig.~\ref{fig:L}\,(b)], with $l(t)=e^{\pi \gamma}e^{-i \gamma \tau}$, where the dimensionless decay rate is given by $\gamma=\vert\mathrm{Im}(\nu)\vert/2$ and $\tau\equiv \Omega t$.

We can gain more insight into the behavior of the Loschmidt echo by studying the trajectories of a classical particle in a modulated harmonic trap---a strategy often employed in semiclassical methods. The connection between the TG gas in a periodically modulated trap and its classical counterpart can be established by noting that the scaling function $\lambda(t)$ which governs the dynamics of the quantum observables in our system 
can be constructed by combining two independent solutions, $\lambda_1(t)$ and $\lambda_2(t)$ to the homogeneous version of the same equation, i.e., Eq.~\eqref{ermakov} with the right hand side equal to zero~\cite{atas2019finite}. The homogeneous equation describes the motion of a classical particle in a periodically modulated harmonic trap, and hence the scaling function $\lambda(t)$ is constructed from purely classical variables or trajectories. 

\begin{figure}[t!]
\includegraphics[width=7.6cm]{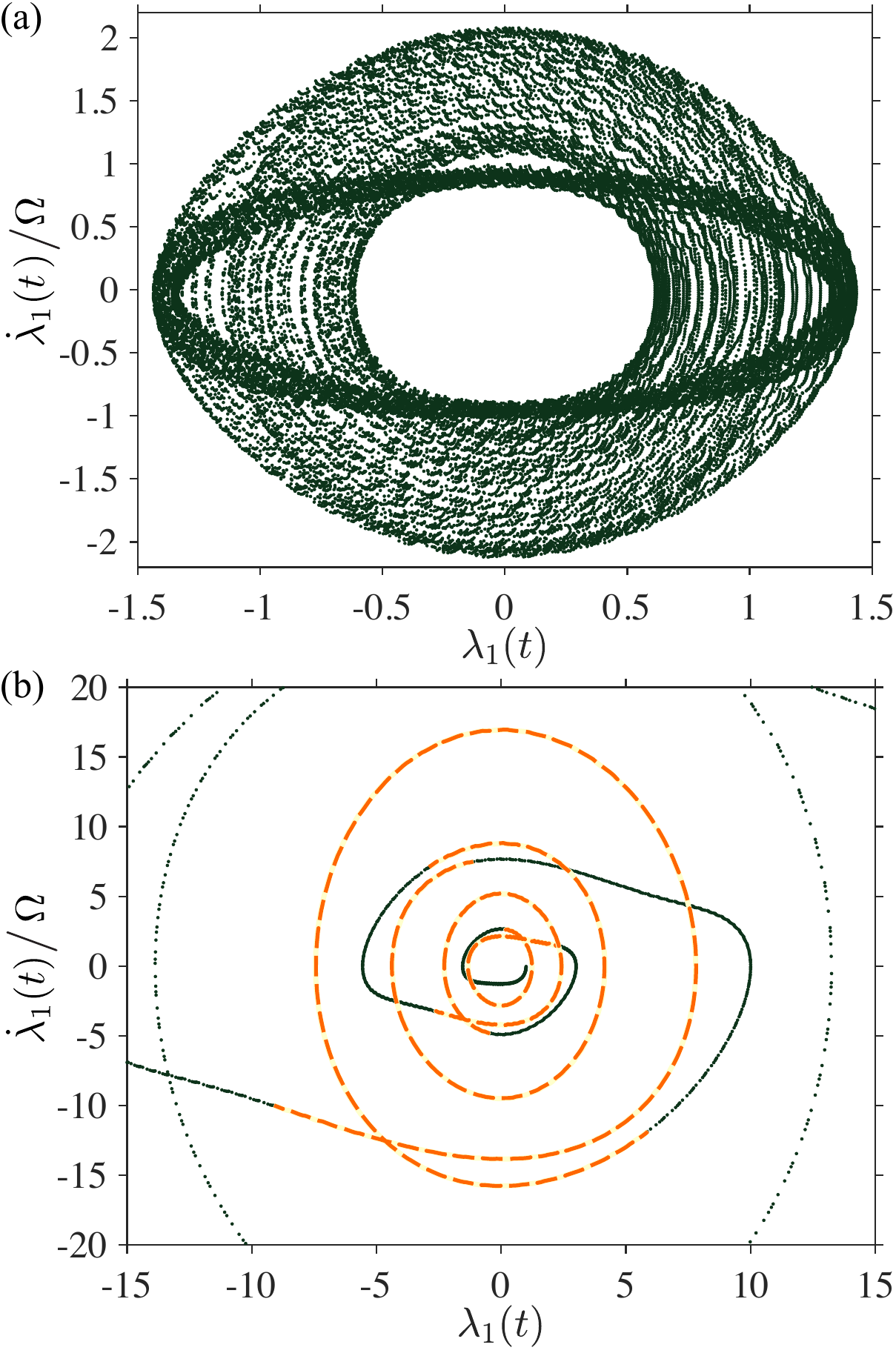}
\caption{The Poincare maps for stable (a) and unstable (b) dynamics. The parameter values are the same as in Fig.~\ref{fig:L}\,(a), \textit{i.e.}, $\alpha=0.9$ for both panels, whereas $a=7.5$ for pane (a), and $a=11$ for panel (b). See text for for further details. 
}
\label{fig:P}
\end{figure}

We use Poincar\'e maps to study whether the characteristic features observed in the behavior of quantum observables, such as the Loschmidt echo, manifest in the behavior of classical trajectories. These maps correspond to sampling--- at regular time intervals---one of the solutions to the classical equation of motion, say $\lambda_1(t)$, in the phases space corresponding to $\lambda_1(t)$ and $\dot{\lambda}_1(t)$. In Fig.~\ref{fig:P}\,(a) and (b)  we plot such Poincar\'e maps representing examples of stable and unstable dynamics,  respectively, for the same parameters as those chosen in Fig.~\ref{fig:L}\,(b).

In the stable regime (Fig.~\ref{fig:P}\,(a)) the classical trajectories are confined to a finite torus.  Thus, in accordance to the Poincar\'e recurrence theorem, if we allow for a sufficiently long evolution time, every point on the torus will be visited at least once. This means that the starting point of the dynamics, $\lambda_1(0)=1$ and $\dot{\lambda}_1(0)=0$ (or its vicinity) is inevitably revisited after some time $t$, causing the scaling solution to return to its initial value, which corresponds to a revival in the Loschmidt echo. 
We emphasize, however, that this is an oversimplified picture because the general scaling solution $\lambda(t)$ is constructed out of two independent solutions of the classical equation of motion, represented by two tori in the phase space. Consequently, an exact
revival happens if a synchronous return to the two different respective initial points in the phase space occurs. 
Furthermore, since the trajectories in the stable regime are confined to finite tori, this means that the Loschmidt echo does not decay to zero in this regime and has a finite non zero lower-bound value. 

In the unstable regime (Fig.~\ref{fig:P}(b)), on the other hand, the classical trajectories spiral out (the support of the trajectories is unbounded) and increase in magnitude, which prevents the Loschmidt echo from revivals or quasi-periodic behavior and explains its overall decay. However, there is a transient dynamics involved here:  the trajectories get trapped in a circular motion for a certain period time and then jump again into the outward spiral motion. The temporary circular motion corresponds exactly to the oscillating plateaux observed in the Loschmidt echo in the decaying example of Fig.~\ref{fig:L}. In Fig.~\ref{fig:P}(b) we have highlighted the nearly circular portions of the trajectory corresponding to the plateaux regions in the Loschmidt echo as dashed orange lines.

\subsection{Quantum Jarzynski equality}

The determinantal approach to evaluating the characteristic function of work distribution can also be used to explicitly verify the quantum Jarzynski equality
\cite{Jarzynski_1997,Kurchan,Esposito_2009,Talkner_2012} in the harmonically trapped TG gas with time-varying trap frequency $\omega(t)$. The Jarzynski equality relates the change in the free energy of the system to the irreversible work along an ensemble of trajectories connecting the initial and final states of the system. To this end we start from the characteristic function $G_{\beta}(\vartheta)$ for a thermal initial state. Making the substitution $\vartheta=i\hbar \beta $ in Eq.~(\ref{xi_eq}) of Appendix \ref{G} leads to $\xi(\vartheta=i\hbar \beta)=e^{-\beta \hbar \omega(t)}$, and therefore the expectation value $\langle e^{-\beta W} \rangle$, entering the quantum Jarzynski equality, can be explicitly evaluated as
\begin{equation}
\langle e^{-\beta W} \rangle \equiv G_{\beta}(\vartheta=i\hbar \beta)=\prod_{i=0}^{\infty} \frac{1+ze^{-\beta \hbar \omega(t)(i+1/2)}}{1+z e^{-\beta \hbar \omega(0)(i+1/2)}}. 
\label{Jarzynski_lhs}
\end{equation}

By inspecting the right hand side of Eq.~\eqref{Jarzynski_lhs}, we see that it is equivalent to the ratio of an instantaneous partition function $\overline{\mathcal{Z}}$ of the system at an effective equilibrium temperature $\beta$ and chemical potential $\mu$,
in a trap of frequency $\omega(t)$, and the actual partition function $\mathcal{Z}_{0}$ of the initial thermal equilibrium state of the system, in a trap of frequency $\omega(0)$. Therefore, Eq.~\eqref{Jarzynski_lhs} can be rewritten as
\begin{equation}
 \langle e^{-\beta W} \rangle =\frac{\overline{\mathcal{Z}}}{\mathcal{Z}_{0}}\equiv e^{-\beta (\overline{\Omega}-\Omega_0)}
 \label{Jarzynski-not}
\end{equation}
where $\overline{\Omega}=-(1/\beta) \log \overline{\mathcal{Z}}$ and $\Omega_{0}=-(1/\beta) \log \mathcal{Z}_{0}$ denote the grand (or Landau) potentials of the non-equilibrium and the initial equilibrium states of the system. The grand potentials here are given by $\overline{\Omega}=\overline{F}-\mu \langle N \rangle$ and $\Omega_0=F_0-\mu \langle N \rangle$, where $\overline{F}$ and $F_0$ are the respective Helmholtz free energies and we have used the fact that our system evolves in the absence of heat or particle exchange, resulting in the convservation of particle number is in each realization of the ensemble. 
Hence we can simplify Eq.~\eqref{Jarzynski-not} to find the standard form of the aforementioned Jarzynski equality given by \cite{Jarzynski_1997,Talkner_2012}
\begin{equation}
 \langle e^{-\beta W} \rangle = e^{-\beta (\overline{F}-F_0)}.
 \label{Jarzynski}
\end{equation}



\section{Summary}

In this work we studied the dynamics of various thermodynamic quantities of the Tonks-Girardeau gas in a time-varying potential $V(x,t)$. Our choice of the TG gas as the working fluid was motivated by the immediate experimental realizability of our proposal this paradigmatic model of a strongly interacting quantum many-body system. Throughout the paper we considered a driving protocol whereby the TG gas started in thermal equilibrium with a reservoir, which is reminiscent of a single stroke of a quantum heat engine or refrigerator cycle. Subsequently it was isolated from the reservoir and was driven out of equilibrium by $V(x,t)$ for a time $t$. Our main result utilized the determinant representation of the many-body wave function for deriving the characteristic function of the quantum work probability distribution and all its moments in terms of the eigenvalues of Fredholm integral operators associated exclusively with single-particle wave functions. Our results, which make no assumptions about the shape of the trapping potential or the functional form of the modulation, allow for numerically efficient evaluation of many-body thermodynamic quantities without the need to construct the full many-body thermal state. 

As an immediate application of our general approach, we considered a TG gas in a harmonic trap, with an arbitrary time variation of its frequency, and presented explicit formulas for all the above thermodynamic quantities, as well as for the nonadiabaticity parameter and the Loschmidt echo. In addition, we have validated, through an explicit analytic derivation, the quantum Jarzynski equality.

We next elaborated on these results for a specific dynamical protocol corresponding to a sinusoidal modulation of the harmonic trap strength with time. This is an experimentally relevant scenario, and our analytic results lend themselves to easy exploration of the parameter space defined by the amplitude and the frequency of the modulation, as well as to gaining fundamental insights into nonequilibrium quantum thermodynamics and its potential quantum technology applications. Furthermore the TG gas in a modulated harmonic trap can display stable or unstable dynamics at a given modulation frequency depending on the amplitude of the modulation. Our results indicate that unstable dynamics, in which the system displays the phenomenon of parametric resonance, has profound effect on all thermodynamic quantities. 
In the case of average work, this corresponds to our ability to drive the system very far from equilibrium and hence to perform a large amount of work on the system in a given amount of time. Even if the system is initialized in its zero-temperature ground state, the unstable dynamics can be observed by the rapid decay of the Loschmidt echo amplitude, which characterizes the sensitivity of the system to the driving protocol and the ergodicity of the semiclassical dynamics of the system. 

{As the determinantal structure of the many-body wave function for the TG gas stems from the Bose-Fermi mapping to the Slater determinant of free fermions, our results are equally applicable to the ideal Fermi gas itself, and for a similar mapping reasons---to a 1D gas of hard-core anyons. As a further extension, our approach can be also applied to any interacting fermionic many-body system that is treated within the self-consistent Hartree-Fock approximation; indeed, the corresponding many-body wave function in this approximation is also a Slater determinant by construction.

Furthermore, even though the examples of thermodynamic quantities that we calculated here were for a harmonically trapped TG gas that possesses scaling solutions, our approach is not exclusively restricted to harmonic traps, but can be applied to other scale-invariant systems. Moreover, the scale invariance itself is not required in general, and instead our approach and the general general results of Sec.~\ref{sec:model}, such as Eqs.~\eqref{eq:Gcompact}--\eqref{Overlap_Matrix}, rely only on the determinantal structure of the many-body wave function. As we explained in the main text after Eq.~\eqref{eigenvalue_expansion_G}, the characteristic function in Eq.~\eqref{eq:Gcompact} can be computed efficiently using a tabulation of the kernels followed by a Nystrom quadrature technique. This approach does not require obtaining the eigenvalues of the Fredholm integral operators. The main reason that the determinants were written in terms of the Fredholm eigenvalues was for convenience and analitical transparency; when applying these results to the harmonically trapped gas, the eigenvalues can be calculated explicitly, hence the utility of Eq.~\eqref{eigenvalue_expansion_G}. However, such exact analytic results for the eigenvalues are not generally available and one has to instead rely on numerical techniques as described above.}

\begin{acknowledgments}
This work was supported through Australian Research Council Discovery Project Grants No. DP170101423 and No. DP190101515.
 \end{acknowledgments}


\appendix 
\section{Characteristic function in terms of Fredholm determinants}
\label{F}

Here we provide the derivation of Eq.~\eqref{eq:Gcompact} of the main text that expresses the characteristic function of the work distribution as a ratio of two Fredholm determinants. Inserting  Eq.~\eqref{work_distribution} into  Eq.~\eqref{Fourier_Transform_Work}, we can rewrite the characteristic function as 
\begin{align}
\nonumber G_{\beta}(\vartheta)&=\frac{1}{\mathcal{Z}_{0}}\!\sum_{N}\!\sum_{s, s^{\prime} } p_{s^{\prime}|s}^{(t)}\, e^{-\beta(E_{N,s}^{(0)}-\mu N)+i\vartheta(E^{(t)}_{N,s^{\prime}}-E_{N,s}^{(0)})/\hbar}\\
\nonumber &=\frac{1}{\mathcal{Z}_{0}}\sum_{N=0}^\infty\frac{1}{(N!)^{2}}\sum_{s_1, \dots, s_N}\sum_{ s_1^\prime, \dots, s_N^\prime} \\
&\hskip 50pt p_{s^{\prime}|s}^{(t)}\, e^{-\beta(E_{N,s}^{(0)}-\mu N)+i\vartheta(E^{(t)}_{N,s^{\prime}}-E_{N,s}^{(0)})/\hbar}, \label{GBappendix}
\end{align}
where the conditional probability reads 
\begin{align}
\notag p_{s^{\prime}|s}^{(t)}=\int \prod_{i=1}^{N}& dx_{i}dx^{\prime}_{i} \Psi_{s}(x_{1},\dots,x_{N},t)  \Phi^{(t)\ast}_{s^{\prime}}(x_{1},\dots,x_{N}) \\
& \times \Psi_{s}^{\ast}(x_{1}^{\prime},\dots,x_{N}^{\prime},t)  \Phi^{(t)}_{s^{\prime}}(x_{1}^{\prime},\dots,x_{N}^{\prime}). \label{conditional_prob}
\end{align}
In Eq.~\eqref{GBappendix}, the sums over the $N$-particle configurations $s$ and $s^{\prime}$ have been explicitly rewritten  as $2N$ independent sums over the single-particle quantum numbers $s_{1}, s_{2},\dots, s_{N}$ and $s_{1}^{\prime}, s_{2}^{\prime},\dots, s_{N}^{\prime}$. The many-body eigenstate energies appearing in the exponential of Eq.~\eqref{GBappendix} can be expressed in terms of the single-particle  eigenenergies as   
\begin{equation}
    E_{N,s}^{(0)}=\sum_{i=1}^{N}E_{s_{i}}^{(0)}, 
\end{equation}
and 
\begin{equation}
    E_{N,s^{\prime}}^{(t)}=\sum_{i=1}^{N}E_{s_{i}^{\prime}}^{(t)}. 
\end{equation}
For a harmonically trapped TG gas, to be treated later, the single-particle energies are given explicitly by $E_{s_{i}}^{(0)}=\hbar \omega(0)\left( s_{i}+\frac{1}{2}\right)$ and $E_{s_{i}^{\prime}}^{(t)}=\hbar \omega(t)\left( s_{i}^{\prime}+\frac{1}{2}\right)$, where $s_i=0,1,2,...$ and $s_i^{\prime}=0,1,2,...$.

We now use the Bose-Fermi mapping, Eq.~\eqref{FB-mapping}, and the Slater determinant form of the fermionic many-body wave functions to write the bosonic wave functions as
\begin{equation}
    \Psi_{s}(x_{1},\dots,x_{N},t)=\frac{A(x_{1},...,x_{N})}{\sqrt{N!}}\,\underset{1\leq n\leq m\leq N}{ \mathrm{det}} \phi_{s_{n}}(x_{m},t),
\end{equation}
and
\begin{equation}
    \Phi^{(t)}_{s^{\prime}}(x_{1},\dots,x_{N})=\frac{A(x_{1},...,x_{N})}{\sqrt{N!}}\,\underset{1\leq n\leq m\leq N}{ \mathrm{det}} \phi_{s_{n}^{\prime}}^{(t)}(x_{m}),
\end{equation}
where $A(x_1,...,x_N)$ is the unit antisymmetric function from Eq.~\eqref{FB-mapping}, whereas
where $\phi_{s_n}(x,t)$ and $\phi_{s_{n}^{\prime}}^{(t)}(x)$ are the time-evolved and the instantaneous single-particle eigenfunctions of the Hamiltonian, respectively. Inserting these expressions back into  Eq.~\eqref{conditional_prob} and interchanging the double integral with the summations over $s$ and $s^{\prime}$, we can rewrite the characteristic function as 
\begin{align}
\nonumber G_{\beta}(\vartheta)=\frac{1}{\mathcal{Z}_{0}}\sum_{N=0}^\infty\frac{1}{(N!)^{2}}\int \prod_{i=1}^{N} dx_{i}dx^{\prime}_{i}& \left(\sum_{s} \Xi_{s}(\mathbf{x},\mathbf{x^{\prime}})\right) \\ \times &\left(\sum_{s^{\prime}} \tilde{\Xi}_{s^{\prime}}(\mathbf{x},\mathbf{x^{\prime}})\right), \label{GBwithintegral}
\end{align} 
where  
\begin{align}
  \notag  \Xi_{s}(\mathbf{x},\mathbf{x^{\prime}})=\frac{1}{N!} \left(\prod_{i=1}^{N}\rho_{s_i}\right) &\,\underset{1\leq n\leq m\leq N}{ \mathrm{det}} \phi_{s_{n}}(x_{m},t)\\  \times & \underset{1\leq p\leq q\leq N}{ \mathrm{det}} \phi_{s_{p}}^{\ast}(x_{q}^{\prime},t), \label{Xis}
\end{align}
and 
\begin{align}
  \notag  \tilde{\Xi}_{s^{\prime}}(\mathbf{x},\mathbf{x^{\prime}})=\frac{1}{N!} \left(\prod_{i=1}^{N}\tilde{\rho}_{s^{\prime}_i}\right) &\,\underset{1\leq n\leq m\leq N}{ \mathrm{det}} \left(\phi_{s_{n}^{\prime}}^{(t)}(x_{m})\right)^{\ast}\\ \times & \underset{1\leq p\leq q\leq N}{ \mathrm{det}} \phi_{s_{p}^{\prime}}^{(t)}(x_{q}^{\prime}).\label{Xisprime}
\end{align}

In Eqs.~(\ref{Xis}) and (\ref{Xisprime}) we have used the shorthand notation $\mathbf{x}=(x_{1},x_{2},\dots, x_{N})$, $\mathbf{x}^{\prime}=(x_{1}^{\prime},x_{2}^{\prime},\dots, x_{N}^{\prime})$ and introduced the parameters 
\begin{equation}
    \rho_{s_i}=\sqrt{z}\exp\left(-E_{s_i}^{(0)}(\beta +i \vartheta/\hbar) \right), \label{rho}
\end{equation}
and 
\begin{equation}
    \tilde{\rho}_{s_i}=\sqrt{z}\exp\left(i \vartheta E_{s_i}^{(t)}/\hbar \right),
    \label{tilde-rho}
\end{equation}
that depend only on the instantaneous single-particle energies $E_{s_i}^{(t)}$, the inverse temperature $\beta$, and the fugacity $z=e^{\beta \mu}$ of the system.

The sums over functions $\Xi_{s}(\mathbf{x},\mathbf{x^{\prime}})$ and $\tilde{\Xi}_{s^{\prime}}(\mathbf{x},\mathbf{x^{\prime}})$ appearing in the integrand of  Eq.~(\ref{GBwithintegral}) have the same form and involve products of two determinants. Let us now show that each of these sums can be further simplified and written in terms of a single determinant. We illustrate the derivation with $\sum_{s^{\prime}}\tilde{\Xi}_{s^{\prime}}(\mathbf{x},\mathbf{x^{\prime}})$, and apply the Leibniz expansion to each of the two determinants in Eq.~\eqref{Xisprime},  denoting all permutations of the set $(1,2,\dots,N)$ for each determinant via $\sigma$ and $\bar{\sigma}$, respectively. Interchanging next the summation over $s^{\prime}=\{s_1',s_2',...,s_N'\}$ with the Leibniz sums over permutations $\sigma$ and $\bar{\sigma}$, we obtain
\begin{align}
\notag    \sum_{s'} &\tilde{\Xi}_{s^{\prime}}(\mathbf{x},\mathbf{x^{\prime}}) \\
\notag
&=\frac{1}{N!}\sum_{\sigma,\bar{\sigma}}(-1)^{\sigma+\bar{\sigma}}\sum_{s^{\prime}_{1}, \dots, s^{\prime}_{N}}  \tilde{\rho}_{s_{1}^{\prime}}\cdots \tilde{\rho}_{s_{N}^{\prime}} \\ &\times \prod_{i=1}^{N}(\phi_{s_{i}^{\prime}}^{(t)}(x_{\sigma_i}))^{\ast} \prod_{j=1}^{N}\phi_{s_{j}^{\prime}}^{(t)}(x_{\bar{\sigma}_j}^{\prime}). \label{Leibnizexpansion}
\end{align}

By regrouping quantities with the same indices and introducing the instantaneous kernel 
\begin{equation}
    g(x,y;t)\equiv \sum_{p}\tilde{\rho}_{p}\phi_{p}^{(t)}(x)(\phi_{p}^{(t)}(y))^{\ast} \label{g_kernel},
\end{equation}
the expansion (\ref{Leibnizexpansion}) can be  simplified to  
\begin{equation}
   \sum_{s'} \tilde{\Xi}_{s^{\prime}}(\mathbf{x},\mathbf{x^{\prime}})= \frac{1}{N!}\sum_{\sigma,\bar{\sigma}}(-1)^{\sigma+\bar{\sigma}}\prod_{i=1}^{N}g(x^{\prime}_{\bar{\sigma}_i},x_{\sigma_i};t),
\end{equation}
which can in turn be recognized as a single determinant $\underset{1\leq n\leq m\leq N}{ \mathrm{det}}g(x_{m}^{\prime},x_{n};t)$. 

Similarly, if we introduce the kernel
\begin{equation}
f(x,y;t)=\sum_{p} \rho_{p}\phi_{p}(x,t)\phi_{p}^{\ast}(y,t), \label{f_kernel}
\end{equation}
then the sum $\sum_{s} {\Xi}_{s}(\mathbf{x},\mathbf{x^{\prime}})$ 
over $s$ in (\ref{GBwithintegral}) is simply given by $\underset{1\leq n\leq m\leq N}{ \mathrm{det}}f(x_{n},x_{m}^{\prime};t)$.

The characteristic function can therefore be written in terms of a product of two determinants in the integrand of Eq.~\eqref{GBwithintegral}:
\begin{align}
\notag G_{\beta}(\vartheta)=&\frac{1}{\mathcal{Z}_{0}}\sum_{N=0}^{\infty} \frac{1}{(N!)^{2}} \int \prod_{i=1}^{N}  dx_{i} dx^{\prime}_{i}  \\
& \times \underset{1\leq n\leq m\leq N}{ \mathrm{det}} f(x_{n},x_{m}^{\prime};t) \underset{1\leq n\leq m\leq N}{ \mathrm{det}} g(x_{m}^{\prime},x_{n};t). \label{G_almost_Fredholm}
\end{align}
Thus, the original expression for the characteristic function, containing a product of four determinants, has been simplified to contain a product of two determinants, one with the kernel (\ref{g_kernel}) and the other with the kernel (\ref{f_kernel}).

We finally make use of the Andr\'eief integration formula \cite{andrief1886note} 
\begin{align}
   \notag & \int \prod_{i=1}^{N}dz_{i}\, \underset{1\leq n\leq m\leq N}{ \mathrm{det}}A_{n}(z_{m})\times \underset{1\leq n\leq m\leq N}{ \mathrm{det}}B_{n}(z_{m}) \\ 
    &=N!\underset{1\leq n\leq m\leq N}{ \mathrm{det}}\left( \int dz \, A_{n}(z)B_{m}(z) \right), \label{andreief_identity}
\end{align}
and eliminate the integral over the primed variable, by taking $z_{i}=x_{i}^{\prime}$, $A_{n}(z_{m})=f(x_{n},x_{m}^{\prime};t)$ and $B_{n}(z_{m})=g(x_{m}^{\prime},x_{n};t)$. This gives an expression for  $G_{\beta}(\vartheta)$ that contains only one determinant,
\begin{equation}
G_{\beta}(\vartheta)=\frac{1}{\mathcal{Z}_{0}}\sum_{N=0}^{\infty} \frac{1}{N!} \int \prod_{i=1}^{N}  dx_{i}  \,\underset{1\leq n\leq m\leq N}{ \mathrm{det}} k(x_{n},x_{m}),  \label{G_Minor_Fredholm_expansion}
\end{equation}
with the kernel 
\begin{align}
k(x,y) &=\int dw f(x,w;t)g(w,y;t) \nonumber \\
       &= \sum_{p,q} \phi_{p}(x,t)k_{pq}(t,\vartheta)\big(\phi_{q}^{(t)}(y)\big)^{*}\label{k_kernel},
\end{align}
where 
\begin{equation}
k_{pq}(t,\vartheta)=\rho_{p}(\vartheta)\tilde{\rho}_{q}(\vartheta)\int dw \phi_{p}^{\ast}(w,t)\phi_{q}^{(t)}(w), \label{Kernel_coefficient}
\end{equation}
and where we have restored the dependence on the variable $\vartheta$ explicitly.

We note that one could have used the Andr\'eief's identity and integrated over the variable $x_{i}$ rather than the primed variable $x_{i}^{\prime}$ as the choice is  arbitrary. Obviously, this alternative should not change the final result for the characteristic function. Carrying out the integration over $x_{i}$, one would obtain the kernel $k^{\ast}(x,y,-\vartheta)$ for the characteristic function, which would reflect the symmetry $G_{\beta}^{\ast}(-\vartheta)=G_{\beta}(\vartheta)$.

Written in the form of Eq.~(\ref{G_Minor_Fredholm_expansion}), the calculation of the characteristic function has so far been reduced to the evaluation of just one determinant involving single-particle wave functions. However,  the evaluation of the multiple integrals still constitutes a challenge such that this form of the characteristic function is still not practical for computation.

In order to further simplify our result for  $G_{\beta}(\vartheta)$ to a more tractable expression, we note that the numerator of Eq.~ (\ref{G_Minor_Fredholm_expansion}) is amenable to a more compact and computational friendly form by recognizing it as the minor expansion of the Fredholm determinant belonging to the kernel $k(x,y)$ \cite{lenard1966one,bornemann2010numerical}. Finally, in order to complete the proof of our main result, Eq.~\eqref{eq:Gcompact}, it remains to show that the grand canonical partition function
\begin{equation}
\mathcal{Z}_{0}\equiv \mathcal{Z}(t=0)= \mathrm{Tr}\left[ e^{-\beta \hat{H}_{0}}\right]=\sum_{N=0}^{\infty}\sum_{s} e^{-\beta(E_{N,s}^{(0)}-\mu N)}, \label{equilibrium_partition_function}
\end{equation}
appearing in the denominator of Eq.~(\ref{G_Minor_Fredholm_expansion}), can also be written as a Fredholm determinant.

In order to show this, we can multiply the exponential factor in Eq.~(\ref{equilibrium_partition_function}) by the identity $1=\int dx_{1}dx_{2}\dots dx_{N}|\Psi_{s}(x_{1},\dots,x_{N},0)|^2$ and expand the many-body wave function using its Slater determinant form.  Interchanging then the multiple sum over $s$  and the  integrals, we find 
\begin{equation}
\mathcal{Z}_{0}=\sum_{N=0}^{\infty}\frac{1}{N!}\int \prod_{i=1}^{N}  dx_{i}  \,\underset{1\leq n\leq m\leq N}{ \mathrm{det}} f_{0}(x_{n},x_{m}). \label{partition_function_minor_expansion}
\end{equation}
Here, the function $f_{0}(x_n,x_m)$ is the equilibrium kernel and corresponds to the function (\ref{f_kernel}) multiplied by $\sqrt{z}$, together with  $t=0$ and $\vartheta=0$; explicitly, it reads as 
\begin{equation}
    f_{0}(x,y)=z\sum_{p=0}^{\infty} e^{-i\beta E_{p}^{(0)} }\phi_{p}(x,0)\phi_{p}^{\ast}(y,0). \label{f0kernel}
\end{equation}
Equation  (\ref{partition_function_minor_expansion}) for the partition function corresponds to the minor expansion of the Fredholm determinant belonging to the kernel $f_{0}(x,y)$.

Collecting  Eqs.~\eqref{G_Minor_Fredholm_expansion} and \eqref{partition_function_minor_expansion} together, we thus arrive at the final  compact  form of the characteristic function in terms of the Fredholm determinants, Eq.~\eqref{eq:Gcompact},
\begin{equation}
    G_{\beta}(\vartheta)=\frac{\mathrm{det}(1+\hat{K})}{\mathrm{det}(1+\hat{F}_{0})},
    \label{eq:Gcompact-Appendix}
\end{equation}
where $\hat{K}$ and $\hat{F}_{0}$ are Fredholm integral operators with kernels (\ref{k_kernel}) and (\ref{f0kernel}), respectively.

Using the expansion of the two determinants in Eq.~\eqref{eq:Gcompact-Appendix} in terms of products over the respective eigenvalues, the characteristic function can be rewritten as
\begin{equation}
G_{\beta}(\vartheta)=\prod_{i}\left( \frac{1+\Lambda_{i}^{(\hat{K})}}{1+\Lambda_{i}^{(\hat{F}_0)}}\right),\label{eigenvalue_expansion_G_Appendix}
\end{equation}
which is Eq.~\eqref{eigenvalue_expansion_G}.

\section{Numerical approach to calculating the eigenvalues of integral operators} 
\label{AppendixB}

In the general case of a TG gas in an arbitrary trapping potential, finding the eigenvalues $\Lambda_{i}^{(\hat{K})}$ of the integral operator with the kernel $k(x,y)$ of the form \eqref{k_kernel}, which is required for evaluating the characteristic function \eqref{eigenvalue_expansion_G_Appendix}, has to rely on numerical approaches. (For a harmonically trapped TG gas, on the other hand, it possible to find analytic solution for this problem, and it will be presented in Appendixes \ref{G} and  \ref{Gaussian kernel eigenvalues}). 

In this appendix, we outline an efficient numerical approach {(different from Nystr\"om direct tabulation of the kernel)} for evaluating the determinant of the Fredholm integral equation  with the kernel $k(x,y)$, for situations when  the single-particle eigenfunctions are not known analytically and instead have to be found numerically as solutions to single-particle Schr\"{o}dinger equation. For this, we write the Fredholm integral equation for the kernel $k(x,y)$, Eq.~\eqref{k_kernel},
\begin{equation}
    \int dy \, k(x,y) \theta^{(\hat{K})}_{i}(y)= \Lambda^{(\hat{K})}_{i}\theta^{(\hat{K})}_{i}(x), 
\end{equation}

By introducing the quantities 
\begin{equation}
    A_{q}^{(i)}=\int dy\, \theta^{(\hat{K})}_{i}(y)\left(\phi_{q}^{(t)}(y)\right)^{\ast},
\end{equation}
we can rewrite the Fredholm integral equation as
\begin{equation}
    \sum_{p,q}\phi_{p}(x,t)k_{pq}A_{q}^{(i)}=\Lambda^{(\hat{K})}_{i}\theta^{(\hat{K})}_{i}(x).    \label{F-matrix}
\end{equation}

We next multiply both side of Eq.~\eqref{F-matrix} by $(\phi_{m}^{(t)}(x))^{\ast}$ and integrate with respect to $x$, yielding
\begin{equation}
    \sum_{p,q} c_{mp}k_{pq}A_{q}^{(i)}=\Lambda^{(\hat{K})}_{i}A_{m}^{(i)}. \label{Fredholm_discrete}
\end{equation}
Here, the coefficients $c_{mp}$ are the coefficients of expansion of the time-evolved eigenstate  $\phi_{p}(x,t)$ in the basis of instantaneous eigenstates, $\phi_{p}(x,t)=\sum_{m}c_{mp}\phi_{m}^{(t)}(x)$, and are given by
\begin{equation}
    c_{mp}=\int dx \, \phi_{p}(x,t)\left( \phi_{m}^{(t)}(x)\right)^{\ast},
\end{equation}
whereas the kernel coefficients $k_{pq}$ are given by Eq.~ \eqref{Kernel_coefficient}.

Let now $\mathbf{C}$ and $\mathbf{K}$ be the matrices with elements $(\mathbf{C})_{ij}=c_{ij}$ and $(\mathbf{K})_{ij}=k_{ij}$ respectively, and $\mathbf{A}^{(i)}$ the vector with elements $A_{q}^{(i)}$. Then Eq.~(\ref{Fredholm_discrete}) can be rewritten in the matrix form,
\begin{equation}
    \left(\mathbf{CK}\right)\mathbf{A}^{(i)}=\Lambda^{(\hat{K})}_{i}\mathbf{A}^{(i)}.
\end{equation}

This last equation shows that the eigenvalues $\Lambda^{(\hat{K})}_{i}$ of the Fredholm integral operator with kernel $k(x,y)$ are the same as the eigenvalues  of the matrix $\mathbf{CK}$, and thus it can now be used for numerical implementation of our eigenvalue problem. In practice, one first calculates the matrix $\mathbf{C}$ of the overlaps between the time-evolved and instantaneous eigenstates. The elements of the matrix $\mathbf{K}$ are then easily obtained through the relation $(\mathbf{K})_{ij}=\rho_{i}\tilde{\rho}_{j}c_{ji}^{\ast}$, where $\rho_i$ and $\tilde{\rho}_j$ are given by Eqs~\eqref{rho} and \eqref{tilde-rho}. The matrix $\mathbf{CK}$ is finally constructed and its determinant is evaluated with a sufficiently large size to ensure convergence.

For kernels of the form $f(x,y)$ and $g(x,y)$, Eqs.~ (\ref{f_kernel}) and (\ref{g_kernel}), the eigenvalue problem is much simpler and the determinant and eigenvalues can be found analytically. Such kernels are known as degenerate kernels in Fredholm theory, and the eigenvalues of the associated Fredholm integral equation are, in fact, given by $\rho_{i}$ and $\tilde{\rho}_{i}$, respectively, without the need for any further calculations.
In order to prove this, one can follow the same argument as the one above for the kernel $k(x,y)$ and show that the matrix equation obtained is already diagonal. This special property is due to the orthogonality of the eigenfunctions appearing in the expansions (\ref{g_kernel}) and (\ref{f_kernel}).

 In particular, for the degenerate kernel $f_0(x,y)$, Eq.~\eqref{f0kernel}, which appears in the determinantal form of the partition function $\mathcal{Z}_0$, Eq.~\eqref{partition_function_minor_expansion}, we observe that it can be expressed in terms of the kernel $f(x,y)$ as $f_{0}(x,y)\equiv \sqrt{z}f(x,y,t=0,\vartheta=0)$. Therefore, its eigenvalues  $\Lambda_{k}^{(\hat{F}_0)}$, required for the evaluation of the denominator of the characteristic function \eqref{eigenvalue_expansion_G_Appendix}, are easily obtained (with the use of Eq.~\eqref{rho}) as
 \begin{equation}
 \Lambda_{k}^{(\hat{F}_0)}=\sqrt{z}\rho_{k}(\vartheta=0)=z\exp(-\beta E_{k}^{(0)}).
 \label{Z-eigenvalue}
 \end{equation}
With this results, we thus arrive at the familiar form of the grand-canonical partition function,
\begin{equation}
    \mathcal{Z}_{0}=\prod_{i=0}^{\infty}(1+\Lambda^{(\hat{F_0})}_i)=\prod_{i=0}^{\infty}(1+ze^{-\beta E_{i}^{(0)}}) \label{partition_function_eigenvalue},
\end{equation}
describing free fermions in one dimension, which is the same as the one for the TG gas due to Bose-Fermi mapping.

\section{Derivation of the characteristic function of a harmonically trapped TG gas for a thermal state}
\label{G}

In this appendix, we show that for a harmonically trapped TG gas the eigenvalues $\Lambda_{i}^{(\hat{K})}$ of the integral operator with the kernel $k(x,y)$, appearing in  the characteristic function \eqref{eigenvalue_expansion_G_Appendix}, can be calculated analytically. Indeed, for a harmonically trapped TG gas, we make use of the Hermite-Gauss polynomials for the time-evolved and instantaneous single-particle wave functions in the expressions  for the kernels $g(x,y)$ and $f(x,y)$, given by Eqs.~(\ref{g_kernel}) and (\ref{f_kernel}). The kernel $k(x,y)$ can then be found from $g(x,y)$ and $f(x,y)$, according to Eq.~\eqref{k_kernel}. 

In order to show this, we first observe that the kernels $g(x,y)$ and $f(x,y)$ have the following generic functional form,
\begin{equation}
    a e^{-b(x^2+y^2)} \sum_{p=0}^{\infty} \frac{H_{p}(c x)H_{p}(c y)}{2^p\, p!}d^p,
\end{equation}
where the constants $a,b,c$ and $d$ (independent of $x$, but dependent on time $t$) are known and depend on the parameters of the single-particle wave functions. Owing to Mehler's summation formula \cite{bateman1953higher},
\begin{align}
 \notag   \sum_{p=0}^{\infty}& \frac{H_{p}(c x)H_{p}(c y)}{2^p\, p!}d^p \\ =&(1-d^2)^{-1/2}
 \exp \left( \frac{2xyd-(x^2+y^2)d^2}{c^2(1-d^2)}\right),
\end{align}
the kernels $g(x,y)$ and $f(x,y)$ can be simplified to a generic form of a Gaussian quadratic in $x$ and $y$ with known coefficients.

The Gaussian quadratic form for the the kernels $g(x,y)$ and $f(x,y)$ allows one to also simplify the expression for the kernel $k(x,y)$, Eq.~\eqref{k_kernel}, which depends on the product of $g(x,y)$ and $f(x,y)$. Specifically, one can integrate the product of the two Gaussian quadratic forms  to obtain another Gaussian quadratic,
\begin{equation}
k(x,y)= A\exp\left( -Bx^2-Cy^2+Dx y\right), \label{Gaussian_kernel}
\end{equation}
where the coefficients $A,B,C$ and $D$ (which are time dependent) are given by
\begin{align}
A&=\displaystyle \frac{z}{\lambda l_{\mathrm{ho}}(0)}\sqrt{\frac{2uv}{\pi\kappa(1-u^2)(1-v^{2})}}, \label{AA} \\
B&\!=\!\displaystyle\frac{1}{2\lambda^{2}l^{2}_{\mathrm{ho}}(0)}\!\left( \frac{1+u^2}{1-u^{2}}\!-\!\frac{4u^{2}\lambda_{ad}^{2}}{\kappa(1-u^{2})^{2}\lambda^{2}}\!-\!i\frac{\dot{\lambda}\lambda}{\omega(0)}\right)\!,
\label{BB}\\
C&=\displaystyle \frac{1}{2l^{2}_{\mathrm{ho}}(0)\lambda_{ad}^{2}}\left( \frac{1+v^2}{1-v^{2}}-\frac{4v^{2}}{\kappa(1-v^{2})^{2}}\right)\!, \label{CC} \\
D&= \displaystyle \frac{4uv}{l^{2}_{\mathrm{ho}}(0) \lambda^{2}\kappa(1-u^{2})(1-v^{2})}.\label{DD}
\end{align}
Here, $l_{\mathrm{ho}}(0)=(\hbar/m\omega(0))^{1/2}$ is the harmonic oscillator length for the initial trap frequency $\omega(0)$, $\lambda(t)$ is the scaling solution, $\lambda_{ad}(t)=(\omega(0)/\omega(t))^{1/2}$ is the scaling solution in the adiabatic limit, $z=e^{\beta \mu}$ is the fugacity, and the quantities $u$, $v$, and $\kappa$ are defined according to
\begin{align}
u&=e^{-\hbar \omega(0)(\beta+i \vartheta /\hbar)}, \\
v&=e^{i \vartheta \omega(t)},\\ 
\kappa&=\left( \frac{1+v^2}{1-v^2}\right)+\left( \frac{\lambda_{ad}}{\lambda}\right)^{2}\left( \frac{1+u^2}{1-u^2}+\frac{i\dot{\lambda}\lambda}{\omega(0)}\right).
\end{align}

We momentarily pause here in order to refer the reader to Appendix \ref{Gaussian kernel eigenvalues}, in which we derive the eigenvalues $\Lambda_{i}^{(\hat{K})}$ of the integral operator with the kernel $k(x,y)$ given in the general form of a Gaussian quadratic \eqref{Gaussian_kernel}. The final result is expressed only in terms of the coefficients $A,B,C$ and $D$ of the quadratic, and reads as [from Eq.~\eqref{Lambda-for-Gaussian}]
\begin{equation}
\Lambda_{i}^{(\hat{K})}=\frac{\sqrt{2\pi}A  D^{i}}{\left[B+C+\sqrt{(B+C)^{2}-D^{2}}\right]^{i+1/2}}. \label{eigenvalue_spectrum}
\end{equation}

Substituting now the expressions for the  coefficients $A,B,C$ and $D$ from Eqs.~\eqref{AA}--\eqref{DD} into Eq.~\eqref{eigenvalue_spectrum}, after a little algebra, we finally obtain
\begin{equation}
\Lambda_{i}^{(\hat{K})}=z \xi^{i+1/2} \label{lambda_K},
\end{equation}
where we have introduced
\begin{widetext}
\begin{equation}
\xi=\frac{4uv}{(1+u^2)(1+v^2)+(1-u^2)(1-v^2)\zeta(t)+\sqrt{\left[(1+u^2)(1+v^2)+(1-u^2)(1-v^2)\zeta(t)\right]^2-16u^2 v^2}},\label{xi_eq}
\end{equation}
\end{widetext}
and where
\begin{equation}
    \zeta(t)=\frac{1}{2\omega(0)\omega(t)}\left( \dot{\lambda}(t)^{2}+\frac{\omega^{2}(0)}{\lambda(t)^2}+\omega^{2}(t)\lambda(t)\right). \label{zeta_SM}
\end{equation}
In the main text, we show that the nonadiabaticity parameter $Q^*(t)$ is equivalent to the above $\zeta(t)$.

Combining  Eq.~\eqref{lambda_K} with the expression \eqref{partition_function_eigenvalue} for the grand-canonical partition function, we can now rewrite Eq.~\eqref{eigenvalue_expansion_G} (or equivalently Eq.~\eqref{eigenvalue_expansion_G_Appendix}) for the characteristic function as
\begin{equation}
    G_{\beta}(\vartheta)=\prod_{i=0}^{\infty}\frac{1+z\xi^{i+1/2}}{1+ze^{-\beta E_{i}^{(0)}}},
\end{equation}
which is  Eq.~\eqref{exact_harmonic_G}.

\section{Eigenspectrum of a Gaussian quadratic kernel}
\label{Gaussian kernel eigenvalues}

As shown in the previous appendix, for a harmonically trapped TG gas, the different kernels (\ref{g_kernel}), (\ref{f_kernel}) and (\ref{k_kernel}) reduce to a Gaussian quadratic form. As an example,  the kernel $k(x,y)$  can be parametrized as
\begin{equation}
k(x,y)= A\exp\left( -Bx^2-Cy^2+Dx y\right),
\label{Gaussian-kernel-gen}
\end{equation}
where $A,B,C$ and $D$ are known coefficients.

In this appendix, we derive an exact expression for the eigenfunctions and eigenvalues of a Fredholm integral equation 
\begin{equation}
\int_{\mathbb{R}}dv ~k(x,y)\theta_{i}^{(\hat{K})}(y)=\Lambda_{i}^{(\hat{K})}\theta_{i}^{(\hat{K})}(x) \label{eigenvalue_equation_app}
\end{equation}
with the Gaussian kernel $k(x,y)$.

We seek the solution for the eigenfunctions  $\theta_{i}^{(\hat{K})}(x)$ in the form of
\begin{equation}
\theta_{i}^{(\hat{K})}(x)= \chi \exp\left( -\eta x^{2}\right)H_{i}\left( \epsilon x \right),
\label{ansatz}
\end{equation}
where $H_{i}(x)$ is  the Hermite polynomial of order $i$, $\chi$ is fixed by the normalization, whereas $\eta$ and $\epsilon$ are constants chosen to satisfy the integral equation (\ref{eigenvalue_equation_app}).

Inserting the ansatz \eqref{ansatz} in the integral equation (\ref{eigenvalue_equation_app}), and making use of the formula $7.374(8)$ from Ref.~\cite{gradshteyn2014table},
\begin{equation}
    \int_{\mathbb{R}}\! e^{-(x-y)^2}\!H_{i}(\alpha x) dx=\sqrt{\pi}(1-\alpha^2)^{i/2}H_{i}\!\left( \frac{\alpha y}{(1-\alpha^2)^{1/2}}\right),
\end{equation}
we find that the left hand side of Eq.~(\ref{eigenvalue_equation_app}) is given by
\begin{align}
\notag \frac{A\chi \sqrt{\pi}}{\sqrt{C+\eta}}&\exp\left(-w^{2}\left(B-\frac{D^2}{4(C+\eta)}\right)\right)\left( 1-\frac{\epsilon^{2}}{(C+\eta)}\right)^{i/2} \\
&\times H_{i}\left( \frac{D\epsilon w}{2\sqrt{(C+\eta)^2-\epsilon^{2}(C+\eta)}}\right).
\end{align}
Therefore, by identification with the right hand side of (\ref{eigenvalue_equation_app}), we deduce the following identities:
\begin{align}
\label{eta_eq}
\eta &=B-\frac{D^{2}}{4(C+\eta)},\\
\label{zeta_eq}
1&=\frac{D}{2\sqrt{(C+\eta)^2-\epsilon^{2}(C+\eta)}},\\
\label{eig-eq}
\Lambda_{i}^{(\hat{K})}&=\frac{A \sqrt{\pi}}{\sqrt{C+\eta}}\left( 1-\frac{\epsilon^{2}}{(C+\eta)}\right)^{i/2}.
\end{align}

Equation \eqref{eig-eq} defines the eigenvalues of the problem in terms of $\epsilon$ and $\eta$. Solving the quadratic equation (\ref{eta_eq}) for $\eta$ gives 
\begin{equation}
\eta=\frac{1}{2}\left(B-C\pm \sqrt{(B+C)^2-D^2}\right),
\end{equation}
where one must choose the positive branch  such that $\mathrm{Re}(\eta)>0$ in order to ensure that the eigenfunctions are normalized. In addition, from Eq.~(\ref{zeta_eq}) we obtain 
\begin{equation}
\epsilon^{2}=(C+\eta)\left( 1-\frac{D^{2}}{4(C+\eta)^{2}}\right),
\end{equation}
which leads to the following final form for the eigenvalues,
\begin{align}
\Lambda_{i}^{(\hat{K})}&=\frac{A\sqrt{\pi}D^{i}}{2^{i}(C+\eta)^{i+1/2}} \nonumber\\ 
&=\frac{\sqrt{2\pi}A  D^{i}}{\left[B+C+\sqrt{(B+C)^{2}-D^{2}}\right]^{i+1/2}},
\label{Lambda-for-Gaussian}
\end{align}
expressed only in terms of the parameters $A,B,C$ and $D$ of the original Gaussian kernel \eqref{Gaussian-kernel-gen}. We note that 
 the eigenvalues are invariant under the exchange $B \leftrightarrow C$, 
 which means that 
 the integration in Eq.~(\ref{eigenvalue_equation_app}) can be done over either the first or second variable of the kernel $k(x,y)$. This in turn means that the kernels $k(x,y)$ and $k(y,x)$ have the same eigenvalue spectrum.

\section{Determinantal form of the Loschmidt echo}
\label{L}
The Loschmidt echo \eqref{L_overlap} is given by the squared overlap between the time-evolved  many-body ground state $\ket{\Psi_{0}(t)}$, and the initial one at time $t=0$, $\ket{\Psi_{0}(0)}$. For $N$ particles in the ground state at $t=0$, the time-evolved and initial states are constructed with the Slater determinant of the corresponding $N$ first single-particle orbitals at time $t$ and time $t=0$, respectively. Explicitly, we have 
\begin{equation}
    \Psi_{0}(x_{1},...,x_{N};t)=\frac{A(x_1,...,x_N)}{\sqrt{N!}}\,\underset{1\leq n\leq m\leq N}{ \mathrm{det}} \phi_{n-1}(x_{m},t), \label{GS_many_body_time_evolved}
\end{equation}
where $A(x_1,...,x_N)$ is the unit antisymmetric function from Eq.~\eqref{FB-mapping}. Similarly, setting $t=0$ in Eq.~(\ref{GS_many_body_time_evolved}), gives the initial ground-state wave function. The overlap between the two wave functions is thus obtained as an $N$-fold integral over the product of two determinants
\begin{align}
    \notag \braket{\Psi_{0}(0)|\Psi_{0}(t)}=\frac{1}{N!}\int &\prod_{i=1}^{N}  dx_{i} \underset{1\leq n\leq m\leq N}{ \mathrm{det}} \phi_{n-1}^{\ast}(x_{m},0) \\& \times \underset{1\leq n\leq m\leq N}{ \mathrm{det}} \phi_{n-1}(x_{m},t).
\end{align}

Owing to Andreief's integration formula (\ref{andreief_identity}), this expression can be further reduced to a form containing just one determinant,
\begin{equation}
\braket{\Psi_{0}(0)|\Psi_{0}(t)}\!=\! \underset{0\leq n\leq m\leq N-1}{ \mathrm{det}}\!\int\! dx\,  \phi_{m}^{\ast}(x,0)\phi_{n}(x,t).
\end{equation}

Introducing a $N\times N$ matrix $\mathbf{A}$, with the matrix elements corresponding to the overlaps between the  time-evolved and initial single-particle wave functions, 
\begin{equation}
    A_{mn}(t)=\int dx\,  \phi_{m}^{\ast}(x,0)\phi_{n}(x,t)
\end{equation}
with $m,n=0,1,\dots,N-1$, one obtains the Loschmidt echo in a simple determinantal form:
\begin{equation}
    L(t)=|\mathrm{det}\, \mathbf{A}|^{2}.
\end{equation}

Evaluating the matrix elements of $\mathbf{A}(t)$ for a harmonically trapped TG gas, with Hermite-Gauss polynomials for the  single-particle eigenfunctions, Eq.~\eqref{Gauss_Hermite_eigenstate}, and the time-evolved wave functions, Eq.~(\ref{Et}), gives
\begin{equation}
A_{mn}(t)=\left( 2^{m+n}m!\,n! \,\pi\lambda(t)\right)^{-1/2}e^{-i\mathcal{E}_n(t)t/\hbar}J_{mn}(t).
\end{equation}
Here, $\mathcal{E}_n(t)$ is given by Eq.~(\ref{Et}), whereas the matrix elements $J_{mn}(t)$ are given by
\begin{equation}
J_{mn}(t)=\int_{-\infty}^{\infty}H_{m}(y)H_{n}\left( \frac{y}{\lambda(t)}\right) e^{-\widetilde{\lambda}(t) y^{2}/2}\,dy,
\end{equation}
where $y=x/l_{ho}(0)$ is the dimensionless coordinate and $\tilde{\lambda}(t)$ is given by Eq.~(\ref{tildelambda}).

Since the parity of the wave function is conserved during the time evolution, transitions between single-particle eigenfunctions with different parity are not allowed. Let us define the $k$-th diagonal of a matrix such that $k=0$ corresponds to the diagonal, $k=\pm 1$ to the upper and lower diagonal respectively and so on. The transition matrix $J_{mn}$ consequently has a special alternating band structure with the $k$-th diagonal equal to zero for odd $k$. Using the generating function method, we find that the matrix elements $J_{mn}(t)$ are given by
\begin{align}
\notag J_{mn}(t)=&\sqrt{\frac{2\pi}{\widetilde{\lambda}}}\bar{n}! i^{\bar{n}} \frac{2^{(m+n)/2}}{\lambda^{n}}\left(\frac{\widetilde{\lambda}-2}{\widetilde{\lambda} \lambda^{2}-2}\right)^{(m-n)/4} \\
&\times \frac{(\widetilde{\lambda} \lambda^{2}-2b^2-2)}{\widetilde{\lambda}^{(m+n)/4}}P_{(m+n)/2}^{|m-n|/2}\Big( \frac{2}{\lambda |\widetilde{\lambda}|}\Big),
\end{align}
where $\bar{n}=\mathrm{min}(m,n)$ and $P_{l}^{q}(z)$ are the associated Legendre polynomials.

At first sight, it does not seem obvious that the matrix $\mathbf{A}(t)$ with such matrix elements will have a simple determinant as to result in the Loschmidt echo amplitude $\mathcal{G}(t)=\mathrm{det}\, \mathbf{A}(t)$ given by Eq~\eqref{L-ampl} and hence the final results for the Loschmidt echo itself, Eq.~\eqref{Loschmidt}, for $N$ particles in the system. However, it is easy to guess the  exact general formula by looking at the determinant of this matrix (of size $N \times  N$) for the first few values of $N$ and one finds the formula (\ref{Loschmidt}). Our result for $\mathcal{G}(t)$ agrees with that of   Ref.~\cite{del2016exact} derived using an alternative approach.

 
%

\end{document}